%% file: acl_latex.tex
\documentclass[11pt]{article}

\usepackage[final]{acl}

\usepackage{times}
\usepackage{latexsym}
\usepackage{amsmath}
\usepackage{amsfonts}
\usepackage[normalem]{ulem}
\usepackage[T1]{fontenc}

\usepackage[utf8]{inputenc}

\usepackage{microtype}

\usepackage{inconsolata}

\usepackage{graphicx}
\usepackage{subcaption}
\usepackage{todonotes}
\usepackage{booktabs}
\usepackage{tabularx}
\usepackage{array}
\usepackage{colortbl}
\usepackage{enumitem}
\usepackage{textcomp}
\usepackage{listings}
\usepackage{fvextra}
\usepackage{ragged2e}
\usepackage{amssymb}
\usepackage{multirow}
\usepackage{tabularray}
\usepackage{makecell}
\usepackage{float}
\usepackage{tcolorbox}
\UseTblrLibrary{booktabs}

\usepackage{xcolor}

\newcolumntype{Y}{>{\raggedright\arraybackslash}X} 
\newcolumntype{C}[1]{>{\centering\arraybackslash}m{#1}} 

\title{
  Do Neural Retrievers Prefer Certain Documents? Evidence of Learned Relevance Priors
}

\author{
  \textbf{Francisco Valentini\textsuperscript{1,3}
  \thanks{Research partially conducted during a stay at the Brno University of Technology, Brno, Czech Republic.}},   
  \textbf{Edgar Altszyler\textsuperscript{2}},
  \textbf{Martin Fajcik\textsuperscript{3}}
  \\
  \textsuperscript{1}CONICET-Universidad de Buenos Aires.\\Instituto de Ciencias de la Computación (ICC). Buenos Aires, Argentina\\
  \textsuperscript{2}Quantit, Buenos Aires, Argentina\\
  \textsuperscript{3}Brno University of Technology, Brno, Czech Republic
  \\
  \small{
    \textbf{Correspondence:} \href{mailto:fvalentini@dc.uba.ar}{fvalentini@dc.uba.ar}
  }
  \\[-4pt]
  \small{
    \textbf{Code:} \texttt{ \url{https://github.com/ftvalentini/RetrievalRelevancePriors}}
  }
}

\begin{document}
\maketitle

\begin{abstract}


Neural retrievers are trained to estimate query-document relevance from annotated query-document pairs. Yet annotation protocols may not purely reflect relevance: they select only a subset of documents for labeling, and this selection can favor certain document types over others. We investigate whether supervised bi-encoder retrievers implicitly learn a document-level relevance prior: a query-independent signal encoded in their representation space as a side effect of training on annotated data. We estimate this prior by training simple classifiers on frozen document embeddings and evaluate three state-of-the-art retrievers across multiple IR benchmarks. We find that supervised neural retrievers encode relevance priors that generalize to unseen documents and are consistent across models. These priors are associated with a findability gap: documents with lower prior are systematically harder to retrieve, even when genuinely relevant. This effect appears in supervised dense retrievers but is weaker and less consistent in BM25, and it persists under controlled matched-document comparisons. Using LLM-based explanations, we find that judged-relevant documents tend to be comprehensive, self-contained summaries of mainstream topics, while niche, fragmentary, or highly technical content is often left unjudged. Retrievers appear to internalize this bias, ranking documents with these favored features higher than documents that lack them, independently of their actual relevance. Our findings expose a structural limitation of supervised retrieval: models trained on annotated data do not just learn relevance, but also the implicit document preferences in their training data.

\end{abstract}

\section{Introduction}

The standard paradigm in information retrieval (IR) trains models to estimate the relevance of a document $d$ to a query $q$: the posterior $\operatorname{P}(R \mid q, d)$, where 
$R$ is a binary or graded indicator of relevance \citep{robertson_probabilistic_2009,lin_pretrained_2021}.
Neural retrievers are trained on datasets of query-document pairs annotated with relevance labels. 
These labels, however, do not purely reflect the notion of relevance: they are shaped by the data collection protocol, which selects only a subset of documents for annotation and leaves many others unjudged. 
If certain document types are systematically labeled as relevant, a retriever may learn to associate relevance not only with query-document compatibility, but also with surface properties of documents, such as topic, style, or format.

More generally, we ask whether supervised neural retrievers implicitly learn a document-level \emph{relevance prior}, $\operatorname{P}(R \mid d)$: a distribution over relevance, independent of any query, reflecting how likely a document is to be deemed relevant based on its content. 
We study this prior as an implicit signal encoded in the representation space of supervised bi-encoders \citep{karpukhin_dense_2020}.
Unlike the explicit document priors in classical IR \citep{berger_information_1999,craswell_relevance_2005},
it is not a design choice, but a side effect of training, learned from the annotation distribution.

Learning such a prior is not inherently problematic.
When a retriever is trained for a specific document collection, the prior can help it focus on likely relevant documents and discard uninformative ones \citep{izacard_memory_2020,fajcik_pruning_2021,chang_neural_2024}.
The concern arises in general-domain retrieval, where the model must generalize to any document collection: 
if training positives are skewed toward certain topics, styles, or formats (as we will show they often are), the retriever may inherit those biases, systematically disadvantaging underrepresented document types at test time.
For instance, a brief practical guide may be harder to retrieve than a comprehensive encyclopedic explanation, even if both are equally relevant to a query.

Our core hypothesis is that supervised retrievers can learn relevance priors reflecting annotation biases in the training data, and that these priors create a \emph{findability} gap: documents with a low relevance prior are harder to retrieve even when they are genuinely relevant.

We organize our investigation around three groups of research questions.
We first ask whether relevance priors emerge at all (\S\ref{sec:priors-general-domain}):
\begin{description}[wide, itemindent=\labelsep, itemsep=\parskip, parsep=0pt, topsep=\parskip]
  \item[\textbf{RQ1.1}] Do supervised bi-encoder retrievers encode relevance priors that generalize to unseen documents and datasets?
  \item[\textbf{RQ1.2}] Are these priors consistent across different retrieval models?
\end{description}

We then ask whether these priors have practical consequences for performance (\S\ref{sec:findability}):
\begin{description}[wide, itemindent=\labelsep, itemsep=\parskip, parsep=0pt, topsep=\parskip]
  \item[\textbf{RQ2.1}] Do relevance priors correlate with document findability, i.e., are documents with lower relevance prior harder to find?
  \item[\textbf{RQ2.2}] Is this effect specific to supervised neural retrievers, or does it also appear in unsupervised lexical methods such as BM25?
  \item[\textbf{RQ2.3}] Does the prior--findability relationship hold controlling for confounding document features?
\end{description}

Finally, we ask \emph{what} drives these priors (\S\ref{sec:drivers}):
\begin{description}[wide, itemindent=\labelsep, itemsep=\parskip, parsep=0pt, topsep=\parskip]
  \item[\textbf{RQ3.1}] How do relevant and unjudged documents differ in content, and how do these differences arise from the annotation process?  
  \item[\textbf{RQ3.2}] What textual features do retrievers learn to associate with relevance?
\end{description}


\section{Problem Formulation} \label{sec:problem-formulation}

Prior work has shown that IR datasets contain systematic document-level biases useful for index pruning (see \S\ref{sec:related-work}). 
Here, we investigate whether retrievers unintentionally learn these biases and how this impacts performance.

Following \citet{berger_information_1999}, we formalize the document-level \emph{relevance prior} as $\operatorname{P}(R \mid d)$: a distribution over relevance for each document $d$. 
For the binary relevance case, the prior probability $\operatorname{P}(R=1 \mid d)$ reflects the likelihood of a document being relevant based on its content alone. 
We focus on priors that are \emph{implicitly} learned: encoded in a retriever's representation space as a side effect of training on annotated data, without any explicit supervision targeting document-level relevance.
Note that $\operatorname{P}(R \mid d)$ is defined by the annotation process, not by a document's true relevance: it reflects the likelihood of a document being \emph{selected} as relevant. Therefore, a low prior score does not imply actual irrelevance.


\subsection{Document Findability}   

A document's \emph{findability}, $F_{\mathcal{R}}(d): D \rightarrow \mathbb{R}$, measures how easily a document $d \in D$ can be retrieved by retriever $\mathcal{R}$ across the queries for which $d$ is relevant, where $D$ is the document collection.
Following \citet{sinha_findability_2023}, we define it as the average rank-based score of $d$ over those queries:

\vspace{-5pt}
\begin{equation}
\label{eq:findability}
F_{\mathcal{R}}(d) = \frac{1}{|Q_d|} \sum_{q \in Q_d} s(p_q, k)
\end{equation}
\vspace{-5pt}

where $Q_d$ is the set of queries for which $d$ is relevant, and $s(p_q, k)$ rewards high rankings $p_q$ of $d$ within the top-$k$ results for query $q$.
Following \citet{sinha_findability_2023}, we use reciprocal rank, so that $s(p_q, k) = 1/p_q$ if $p_q \leq k$, and $0$ otherwise.\footnote{We set $k=100$. A document at rank 1 scores 1; at rank 2, 1/2; at rank 100, 1/100; and beyond rank 100, 0.}


\subsection{Relevance Prior Estimation}

To assess whether $\mathcal{R}$ implicitly encodes a relevance prior, we attempt to recover this signal from its representation space.
We treat estimating $\operatorname{P}(R \mid d)$ as a binary classification problem.
Let $\phi_\mathcal{R}(d) \in \mathbb{R}^k$ be the frozen document embedding produced by $\mathcal{R}$.
Without query information, relevance is modeled as a function of this embedding alone, $\operatorname{P}(R \mid d) = f(\phi_\mathcal{R}(d))$, quantifying how much 
relevance structure 
$\mathcal{R}$ has encoded in its document representations.
We call $f$ the \emph{prior model}.

We instantiate $f$ as a logistic regression classifier trained on positive and negative documents from $\mathcal{R}$'s training set, using $\phi_\mathcal{R}(d)$ as input features: $\hat{P}(R \mid d) = \sigma\!\left(\mathbf{w}^\top \phi_\mathcal{R}(d) + b\right)$.
The degree to which a linear classifier separates positive from negative documents in embedding space measures how accessibly $\mathcal{R}$ has encoded relevance information.
If this signal generalizes to held-out data, it suggests the retriever has learned a document-level relevance prior exploitable at test time: evidence that annotation biases have been internalized into its document representations.

\begin{figure*}
  \centering
  \begin{subfigure}[b]{0.748\textwidth}
    \centering
    \includegraphics[width=\columnwidth]{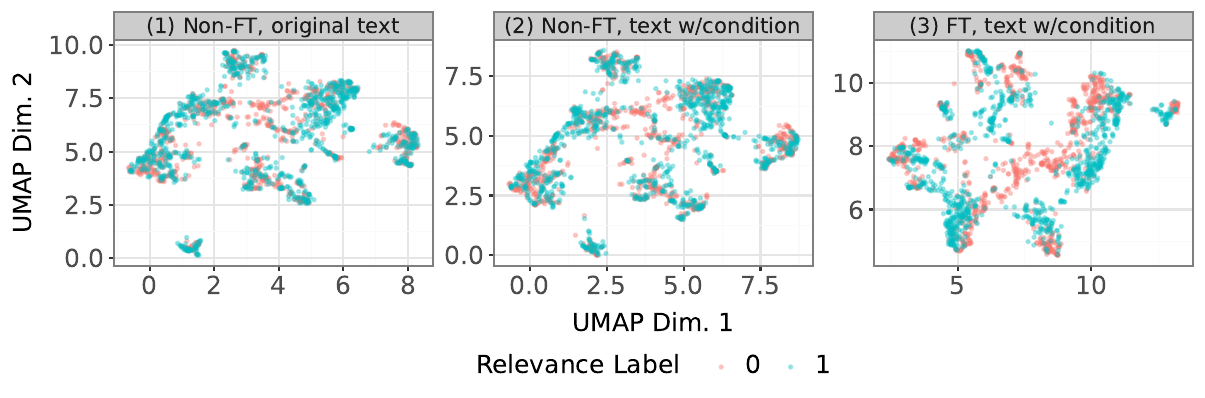}
    \vspace{-18pt}
    \caption{}
    \label{fig:toy-train-umap}
  \end{subfigure}
  \begin{subfigure}[b]{0.246\textwidth}
    \centering
    \includegraphics[width=\columnwidth]{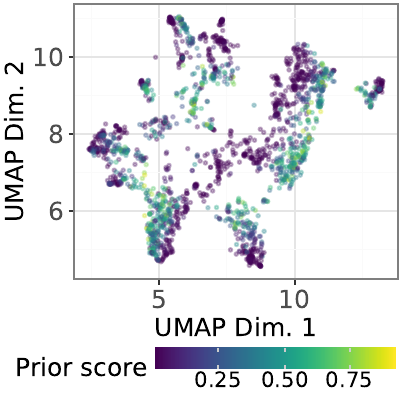}
    \vspace{-18pt}
    \caption{}
    \label{fig:toy-train-umap-prior-score}
  \end{subfigure}
  \vspace{-18pt}
  \caption{
    UMAP projections of E5 document embeddings for a random sample of 1k positive and 1k negative training documents.
    \textbf{(a)} Embeddings under three conditions: the original pre-trained model on original documents, the pre-trained model after injecting the spurious token ($M{=}0.9$), and the model fine-tuned on biased data ($M{=}0.9$).
    \textbf{(b)} The same fine-tuned embeddings colored by relevance prior score.
  }
  \label{fig:toy-train-umap-toy-full}
\end{figure*}

\begin{figure}
  \centering
  \includegraphics[width=1\columnwidth]{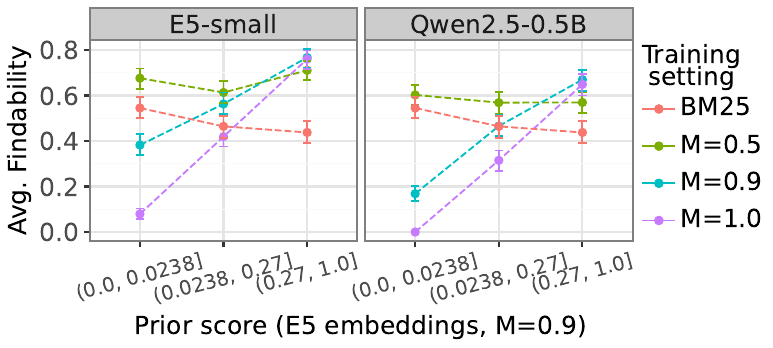}%
  \caption{
    Model findability versus prior scores from E5-small fine-tuned with $M{=}0.9$.
    Error bars show 95\,\% bootstrap confidence intervals over documents within each bin.
    The relationship between the actual prior condition and findability is shown in Fig.~\ref{fig:toy-prior-vs-condition}, App.~\ref{app:toy-experiment}.
  }
  \label{fig:toy-prior-vs-findability}
\end{figure}

\section{Motivating Toy Experiment} \label{sec:toy-experiment}

To test whether retrievers can learn a relevance prior and whether it affects document findability, we injected a known spurious prior into training data and measured its downstream effects.

\paragraph{Setup.}
We built a training set of 1,000 positive query-document pairs from LoTTE \citep{santhanam_colbertv2_2022}, with a document collection consisting of these documents plus $\sim$200k unjudged ones matched in topic and length across the 5 LoTTE topics. 
Positive and unjudged documents were kept as similar as possible so that any learned difference can be attributed to the injected condition alone. 
For each query, hard negatives were mined with BM25 \citep{robertson_okapi_1995} over the collection.

We simulated a relevance prior by prepending string `\texttt{[X]}' to a fraction $M$ of relevant documents and $(1-M)$ of unjudged ones, making it a spurious but reliable relevance signal in training.
We tested $M \in \{1.0, 0.9, 0.5\}$, from fully biased to balanced.
The test set, 1k LoTTE queries, is always balanced: 50\,\% of relevant documents carry \texttt{[X]} and 50\,\% do not, so no prior signal is available at test time.

We fine-tuned two small pre-trained bi-encoders on each setting: E5-small\footnote{\href{https://huggingface.co/intfloat/e5-small-unsupervised}{intfloat/e5-small-unsupervised}} \citep{wang_text_2024} and Qwen2.5-0.5B\footnote{\href{https://huggingface.co/Qwen/Qwen2.5-0.5B}{Qwen/Qwen2.5-0.5B}} \citep{qwen_team_qwen25_2024}.
See App.~\ref{app:toy-experiment} for training details.

\paragraph{Results.}
\uline{Fine-tuning on biased data caused positive and unjudged documents to occupy distinct regions in embedding space; the base model showed no such separation} (Fig.~\ref{fig:toy-train-umap}).
A prior model trained on document embeddings 
tracked this separation, showing that the injected prior can be recovered from the representation space (Fig.~\ref{fig:toy-train-umap-prior-score}).
Stronger annotation bias (higher $M$) produced stronger correlation between prior scores and the spurious condition on held-out data, confirming the prior model captures the degree of bias absorbed during training (Fig.~\ref{fig:toy-prior-vs-condition}, App.~\ref{app:toy-experiment}).

Crucially, \uline{the learned prior affected retrieval: documents lacking the spurious condition, and thus with lower prior score, were harder to retrieve for biased neural models, while BM25 and unbiased models showed no such effect} (Fig.~\ref{fig:toy-prior-vs-findability}).
This also illustrates how a single prior model can measure bias in different retrievers trained on the same data.
By scoring all documents with one reference model (here based on E5-small with $M{=}0.9$), we can compare how sensitive different retrievers are to that prior:
those trained with weaker biases ($M{=}0.5$) show a flat relationship with findability, while those trained with stronger biases ($M{=}0.9$ and $M{=}1.0$) show progressively steeper slopes.

\medskip
Together, these results establish two key points: 
\begin{enumerate}[wide, itemindent=\labelsep, itemsep=\parskip, parsep=0pt, topsep=\parskip]
  \item Neural retrievers \emph{can} encode systematic, query-independent differences between relevant and unjudged documents, forming a relevance prior absent in lexical models like BM25.
  \item This prior \emph{can} create a findability gap: documents with lower relevance prior are harder to retrieve even when they are genuinely relevant.
\end{enumerate}

\begin{figure}
  \centering
  \includegraphics[width=1.0\columnwidth]{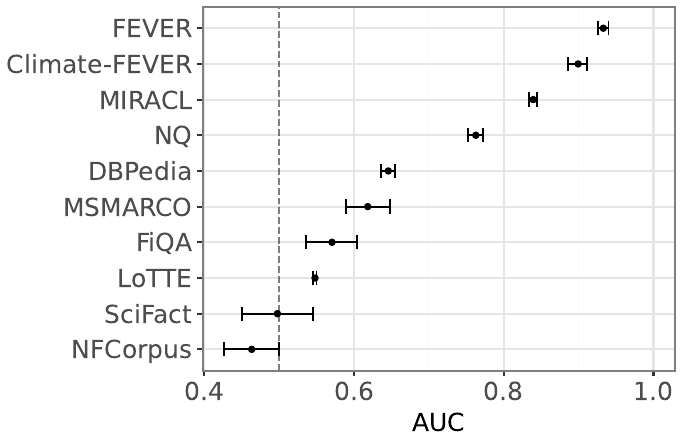}%
  \caption{
    AUC of the \textsc{bge}-based prior model on held-out documents across datasets.
    Error bars show 95\,\% bootstrap confidence intervals.
    Refer to App.~\ref{app:prior-model} for details on dataset counts and AUC values.
  }
  \label{fig:heldout-auc}
\end{figure}

\section{Relevance Priors in General-domain Retrievers} \label{sec:priors-general-domain}

To test whether relevance priors emerge in actual supervised pipelines, we studied three high-performing bi-encoders in the MTEB benchmark \citep{muennighoff_mteb_2023}:
bge-en-icl\footnote{\href{https://huggingface.co/BAAI/bge-en-icl}{BAAI/bge-en-icl}} \citep{li_making_2024},
NV-Embed-v2\footnote{\href{https://huggingface.co/nvidia/NV-Embed-v2}{nvidia/NV-Embed-v2}} \citep{lee_nv-embed_2024},
and gte-Qwen2-7B-instruct\footnote{\href{https://huggingface.co/Alibaba-NLP/gte-Qwen2-7B-instruct}{Alibaba-NLP/gte-Qwen2-7B-instruct}} \citep{li_towards_2023} (hereafter \textsc{bge}, \textsc{nv-embed}, and \textsc{gte}).
We used \textsc{bge} as our anchor model, because its exact training data is publicly available\footnote{\href{https://huggingface.co/datasets/cfli/bge-full-data}{cfli/bge-full-data}}, and trained a prior model on top of its frozen document embeddings.

We estimated $\operatorname{P}(R \mid d)$ with a binary classifier using the full \textsc{bge} train set, which combines retrieval and non-retrieval tasks (e.g., clustering and classification).
Positive examples are relevant documents; negatives are the hard negatives provided in the dataset.
We kept all benchmarks because our goal is to recover the priors learned during fine-tuning, not to isolate retrieval data's contribution (see training data composition in Table~\ref{tab:bge-train-data-composition}, App.~\ref{app:prior-model}).

\subsection{Prior Model Evaluation}

If the retriever has learned a generalizable relevance prior, the prior model should distinguish unseen relevant documents from unjudged ones. 
To test this, we evaluated the model on held-out documents from multiple IR datasets using AUC, treating judged-relevant documents as positives and randomly sampled unjudged documents as negatives, with no overlap with training data (see App.~\ref{app:prior-model} for dataset selection details).
Performance above chance (AUC > 0.5) indicates that systematic differences between relevant and unjudged documents are encoded in the retriever's embeddings and generalize beyond training.

\uline{The retriever encodes a relevance prior that generalizes to unseen documents, though its strength varies across datasets} (Fig.~\ref{fig:heldout-auc}).
High AUC in datasets like FEVER \citep{thorne_fever_2018}, Climate-FEVER \citep{diggelmann_climate-fever_2020}, MIRACL \citep{zhang_miracl_2023}, and Natural Questions \citep{kwiatkowski_natural_2019} suggests strong priors driven by annotation biases.
For instance, FEVER queries (claims) sourced from popular Wikipedia articles likely bias relevant documents toward well-known entities (see \S\ref{sec:drivers}).
Datasets like NFCorpus \citep{boteva_full-text_2016} and SciFact \citep{wadden_fact_2020} yielded near-chance AUC, indicating either no systematic bias, that the embeddings do not encode it, or poor generalization to held-out data.

We stress that separability is a \emph{necessary} but not a \emph{sufficient} condition for a prior learned during fine-tuning.
An above-chance AUC shows that a query-independent relevance signal is linearly accessible in the retriever's document representations; it does not by itself establish that the signal comes from fine-tuning rather than from the pre-trained backbone.
To provide further evidence, App.~\ref{app:backbone-comparison} compares the AUCs of \textsc{bge} with those of its backbone and finds that fine-tuning strengthens separability on most datasets.

\begin{table}
  \centering
  \small
  \input{tables/prior_models_correlations.tex}

  \caption{
    Pearson correlation between prior scores derived from different retrievers, evaluated on a random sample of 10,000 documents from LoTTE and MIRACL.
    All values are statistically significant (p < 0.001).
  }
  \label{tab:prior-models-correlations}

\end{table}

\begin{figure*}[!htbp]
  \centering
  \includegraphics[width=1.0\textwidth]{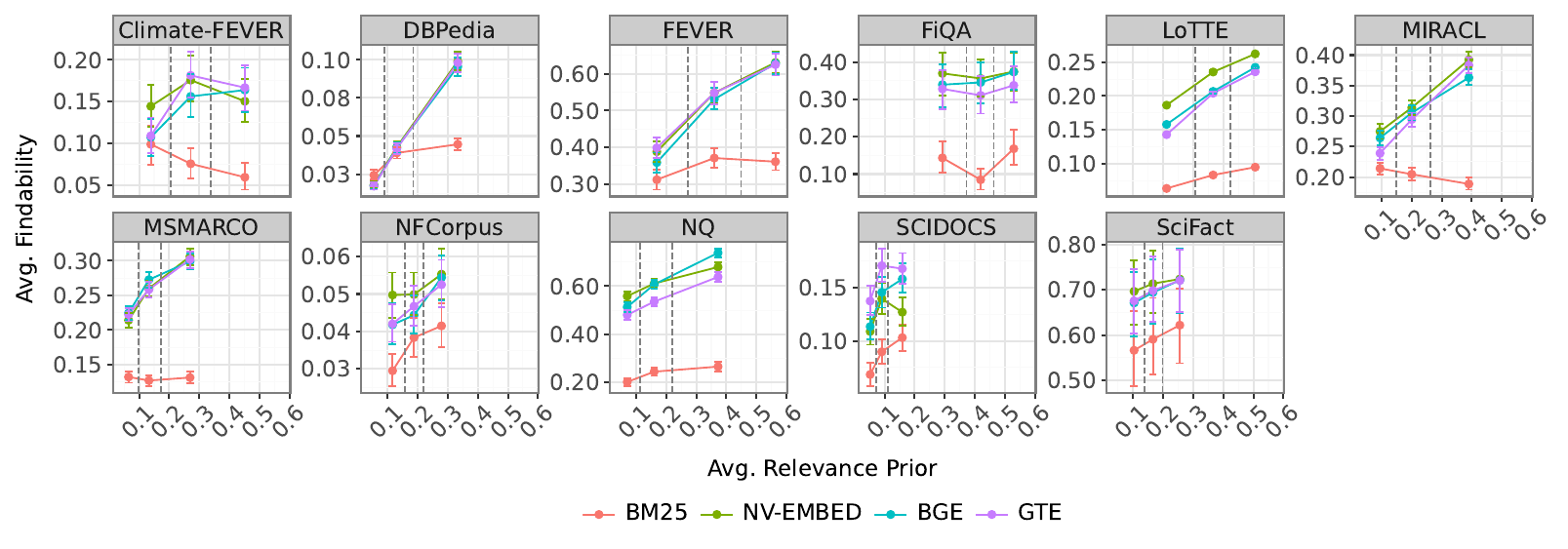}%
  \caption{
    Average findability vs. average relevance prior across datasets.
    Bins contain equal numbers of documents per dataset.
    Error bars show 95\,\% bootstrap confidence intervals over documents within each bin.
  }
  \label{fig:prior-vs-findability}
\end{figure*}

\subsection{Cross-model Consistency} \label{sec:cross-model-consistency}

We next ask whether the priors encoded by \textsc{bge} are model-specific or shared across retrievers trained on similar data. 
We trained separate prior models on embeddings from each retriever using one million randomly sampled training documents, and then computed probabilities for 10,000 random documents per collection from LoTTE (Stack Exchange answers) and MIRACL (Wikipedia paragraphs).
The choice was motivated by their scale, topical diversity, and the fact that both are outside the prior model training data.

Prior scores were moderately to highly correlated across models (0.5--0.7 Pearson $r$), meaning that although absolute scores vary, high-prior documents in one model tend to score highly across the others; that is, \uline{different bi-encoders independently encode similar document priors} (Table~\ref{tab:prior-models-correlations}; see Fig.~\ref{fig:prior-models-correlations}, App.~\ref{app:prior-model} for scatter plots).
Despite differences in base models and training procedures, the significant overlap in training data suggests retrievers share the same underlying annotation biases.

\medskip

These findings affirmatively answer \textbf{RQ1}: 
\begin{description}[wide, itemindent=\labelsep, itemsep=\parskip, parsep=0pt, topsep=\parskip]
  \item[\textbf{RQ1.1}] Supervised retrievers encode relevance priors: when datasets introduce systematic differences between relevant and unjudged documents, bi-encoders capture this signal in their embeddings, which generalizes to new documents.
  \item[\textbf{RQ1.2}] Priors are consistent across retrievers.
\end{description}


\section{The Effect of Priors on Document Findability} \label{sec:findability}

We now explore whether relevance priors have practical implications for retrieval performance.
Our hypothesis is that documents with higher relevance priors are easier to find for supervised neural retrievers, which implicitly learn these priors from training data.
Unsupervised methods like BM25, which rely on lexical statistics rather than relevance judgments, should not show this pattern.

While BM25 does not learn from relevance judgments, it is not prior-free: its scoring function encodes implicit biases of a different nature, such as document length normalization \citep{robertson_probabilistic_2009} and term rarity effects \citep{gu_learning_2016}.
These inductive biases, however, differ from the learned supervised priors that emerge from training on annotated data.

To assess the relationship between findability and relevance prior, we scored each relevant document from our evaluation datasets using our prior model (\S\ref{sec:priors-general-domain}), and computed its findability (Eq.~\ref{eq:findability}) across three dense neural retrievers and BM25, on test and dev queries, excluding documents from the prior model's training data.
We then grouped documents into equal-sized bins by prior score and computed the mean prior and findability within each bin (dataset statistics in Table~\ref{tab:findability-curves-counts}, App.~\ref{app:prior-model}).

\uline{Findability increases with relevance prior for supervised neural retrievers across most datasets} (Fig.~\ref{fig:prior-vs-findability}).
This trend is pronounced in several cases (e.g., DBpedia, MIRACL, FEVER, LoTTE, MSMARCO, Natural Questions) and weaker in others (e.g., Climate-FEVER, FIQA, NFCorpus, SciFact, SCIDOCS), but holds across all three evaluated bi-encoders. 
\uline{BM25, by contrast, shows no consistent trend}: the correlation between prior and findability is sometimes positive (e.g., DBpedia, LoTTE), sometimes negative (e.g., Climate-FEVER, MIRACL), and sometimes flat (e.g., MSMARCO).

\subsection{The Prior Effect Persists Under Matched Comparisons} \label{sec:isolating-prior-effect}

The previous analysis supports our hypothesis: neural retrievers, trained on relevance labels, are implicitly biased toward high-prior documents, whereas BM25 is not.
However, this observed relationship need not be causal; it may reflect confounding factors, of which two are particularly relevant.

The first source is document-level features.
For example, documents rich in named entities may generate more specific queries that are easier to match.
Longer documents may also produce queries with more matching terms, making retrieval easier regardless of any learned prior. 
Indeed, document length correlates moderately with relevance prior across many datasets (Table~\ref{tab:length-vs-prior}, App.~\ref{app:isolating-prior-effect}).
These features could produce an apparent prior--findability relationship even in a model with no learned biases, affecting even BM25.

The second source is dataset-specific characteristics.
Some datasets use graded relevance labels (e.g., MSMARCO, DBpedia-Entity, NFCorpus): more relevant documents may simply be easier to retrieve, and if higher grades correlate with higher priors, this would confound the relationship.
Additionally, a document's expected findability is mechanically influenced by the number of competing relevant documents for the same query.\footnote{Consider a perfect retriever: if a query has ten relevant documents, their expected reciprocal rank will be well below 1 even if all are retrieved at the top; if a query has only one, its reciprocal rank is exactly 1.}

\begin{figure}[!htbp]
  \centering
  \includegraphics[width=1.0\columnwidth]{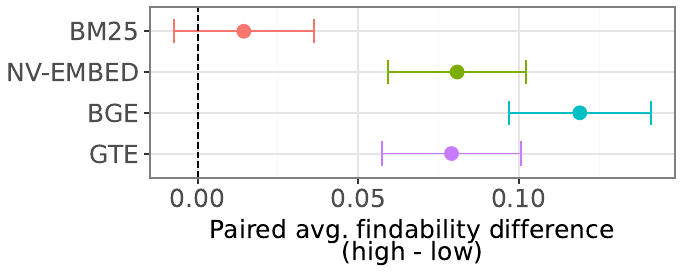}%
  \caption{
    Average findability difference between high- and low-prior documents.
    Error bars show 95\,\% bootstrap confidence intervals.
  }
  \label{fig:matched-findability}
\end{figure}

To isolate the effect of relevance prior from these confounders, we designed a controlled experiment using matched pairs of low- and high-prior documents that are otherwise as similar as possible in terms of their confounders (see App.~\ref{app:isolating-prior-effect}).
Using all Wikipedia paragraphs from MIRACL, we defined low-prior documents as those in the bottom 10\,\% of the prior distribution and high-prior documents as those in the top 10\,\%, and then matched each high-prior document to a similar low-prior one across potentially confounding features.
For each matched pair, we generated synthetic queries with an LLM (see App.~\ref{app:isolating-prior-effect}).
This also lets us evaluate retrieval on very low-prior documents, which are rarely annotated as relevant in benchmarks and thus typically have no natural queries.

After controlling for confounders, supervised neural retrievers find high-prior documents substantially easier to retrieve than low-prior ones, while BM25 shows a considerably smaller gap (Fig.~\ref{fig:matched-findability}).
This \uline{suggests that learned relevance priors themselves contribute to retrieval performance}.


\medskip

We have now answered \textbf{RQ2}:
\begin{description}[wide, itemindent=\labelsep, itemsep=\parskip, parsep=0pt, topsep=\parskip]
  \item[\textbf{RQ2.1}] Learned relevance priors are associated with retrieval performance: documents with lower relevance prior are consistently harder to find for supervised neural retrievers. 
  \item[\textbf{RQ2.2}] In unsupervised lexical methods like BM25, the relationship between relevance prior and findability is weaker and less consistent.
  \item[\textbf{RQ2.3}] Even when controlling for confounders via matched documents, supervised neural retrievers still favor high-prior documents over comparable low-prior ones. BM25 does not, suggesting the prior itself contributes to the findability gap.
\end{description}

\section{The Drivers of Relevance Priors} \label{sec:drivers}

This section addresses two related questions: why relevant and unjudged documents are strongly separable in some IR datasets (\S\ref{sec:drivers-annotation}), and what features make a document low- or high-prior for supervised retrievers (\S\ref{sec:drivers-priors}).
We use a shared methodology for both: LLM-based explanations.

\subsection{LLM-Based Explanations} \label{sec:llm-explanations}

We use a two-stage framework to identify systematic differences between two document groups.
Let $\mathcal{D}_a$ and $\mathcal{D}_b$ be two document collections.
We seek a natural-language explanation $e$ of their differences.
First, we sample $n$ document pairs $\mathcal{P} = \{(d_a^{(i)}, d_b^{(i)})\}_{i=1}^{n}$, where $d_a^{(i)} \sim \mathcal{D}_a$ and $d_b^{(i)} \sim \mathcal{D}_b$.
An LLM $\mathcal{M}_1$ is prompted with $\mathcal{P}$ to produce $e = \mathcal{M}_1(\mathcal{P})$: a natural-language description of the distinguishing features between the two groups.

In the second stage, we assess whether $e$ reflects a generalizable pattern rather than a sampling artifact.
We sample a balanced held-out set $\mathcal{H} = \{(d^{(j)}, y^{(j)})\}_{j=1}^{m}$, where $d^{(j)} \sim \mathcal{D}_{y^{(j)}}$, $d^{(j)} \notin \mathcal{P}$, and $y^{(j)} \in \{a,b\}$ 
denotes group membership, with $m/2$ documents per group.
A second LLM $\mathcal{M}_2$ classifies each held-out document using $e$ as a prompt: $\hat{y}^{(j)} = \mathcal{M}_2\!\left(d^{(j)},\, e\right)$.
We report classification accuracy $\text{Acc}(e) = \frac{1}{m} \sum_{j=1}^{m} \mathbf{1}\!\left[\hat{y}^{(j)} = y^{(j)}\right]$ as a measure of explanation reliability: $e$ captures a stable regularity if $\text{Acc}(e)$ substantially exceeds the random-chance baseline of $0.5$.
Prompts, model settings, and further details are in App.~\ref{app:llm-explanations}.

This approach is inspired by prior work on LLM-based hypothesis generation \citep{singh_explaining_2023,zhong_goal_2023,zhou_hypothesis_2024} evaluated via predictive accuracy \citep{bills_language_2023,huben_sparse_2023}.
Rather than relying on a predefined feature set, this method identifies differences open-endedly, producing human-readable hypotheses about differences between document groups, which can be tested for generalizability.


\subsection{What Do Annotated-Relevant Documents Look Like?} \label{sec:drivers-annotation}

What textual differences distinguish documents labeled as relevant from unjudged ones?
To characterize these annotation biases, we focus on the four datasets with the highest separability in \S\ref{sec:priors-general-domain}.
For each, we sampled $n{=}500$ test document pairs to generate an explanation $e$, validated on $m{=}250$ held-out documents (125 per class).
Below we summarize these explanations and hypothesize connections to each dataset's collection protocol (App.~\ref{app:llm-explanations} includes the full explanations and examples):

\begin{description}[wide, itemindent=\labelsep, itemsep=\parskip, parsep=0pt, topsep=\parskip]
  \item \textbf{Climate-FEVER} (Acc.=0.896). 
  Unjudged documents are short stubs about minor, localized topics; relevant documents are long, detailed overviews of major global concepts and climate science.
  This follows from the dataset design: relevant documents are inherently climate-related, while unjudged ones span all of Wikipedia.
  
  \item \textbf{FEVER} (Acc.=0.808). 
  Unjudged documents are short stubs with details about niche subjects; relevant documents are thorough summaries of notable, mainstream topics.
  This aligns with FEVER queries being generated from highly visited Wikipedia articles, biasing relevant documents toward widely known entities.
  
  \item \textbf{MIRACL} (Acc.=0.672). 
  Unjudged documents are mid- or end-of-article paragraphs with unresolved pronouns and missing context; relevant documents are self-contained lead paragraphs that introduce the subject clearly.
  This reflects the MIRACL annotation process, where queries were written from the first 100 words of Wikipedia articles.

  \item \textbf{Natural Questions} (Acc.=0.640). 
  Unjudged documents are fragmented passages lacking context; relevant documents are coherent introductory paragraphs with clear summaries.
  This is consistent with the predominance of simple factual queries in NQ, which are drawn from real Google searches (e.g., \textit{who/what is X}). 
  Wikipedia introductions are specifically designed to answer such questions.

\end{description}

Across all four datasets, a common pattern emerges:
\uline{unjudged documents tend to concern niche, peripheral, or context-dependent content, while relevant documents provide comprehensive, self-contained summaries of mainstream topics.}
This pattern is strongest in Climate-FEVER and FEVER (Acc.\ $\geq 0.808$), and weaker in MIRACL and NQ (Acc.\ $\leq 0.672$), where document position within an article matters more than topic popularity.

\subsection{What Do High-Prior Documents Look Like?} \label{sec:drivers-priors}

Having examined how annotation protocols shape which documents get labeled relevant, we now ask: what textual features do supervised retrievers internalize as relevance signals? 
Do these match the features that distinguish relevant from unjudged documents?
We derive explanations separately for four collections spanning diverse domains: MIRACL (Wikipedia paragraphs), LoTTE (Stack Exchange answers), MSMARCO (web pages), and SCIDOCS (scientific abstracts).

Since relevance priors are continuous, we formed two groups per dataset using 25th/75th percentile cutoffs.  
As before, we used $n{=}500$ pairs to generate $e$ and validate on $m{=}250$ documents.
Note that absolute prior values at these thresholds vary by dataset, reflecting differences in overall prior distributions (see Fig.~\ref{fig:prior-distribution}, App.~\ref{app:llm-explanations}). 
Below we summarize the generated explanations (App.~\ref{app:llm-explanations} includes the full explanations and examples):

\begin{description}[wide, itemindent=\labelsep, itemsep=\parskip, parsep=0pt, topsep=\parskip]
  \item \textbf{MIRACL} (Acc.=0.776): High-prior paragraphs are comprehensive and explanatory; low-prior documents are brief, fragmentary passages providing isolated facts or structural filler.
  \item \textbf{LoTTE} (Acc.=0.736): High-prior answers are theoretical texts that build conceptual understanding; low-prior ones are practical, step-by-step guides for specific tasks.
  \item \textbf{MSMARCO} (Acc.=0.692): High-prior documents present clean, objective educational content; low-prior documents are poorly formatted, disjointed, or commercially oriented.
  \item \textbf{SCIDOCS} (Acc.=0.592): High-prior abstracts cover health and behavior, use structured abstracts with capitalized headers, and report clinical findings; low-prior abstracts focus on technical, math-heavy fields with single-paragraph abstracts validated through system benchmarks.
\end{description}

The high accuracies for MIRACL, LoTTE, and MSMARCO suggest the LLM identified reliable, generalizable patterns.
The lower accuracy for SCIDOCS suggests differences between high- and low-prior documents are subtler in this domain, or that the explanation or classifier failed to capture them.

Overall, a consistent pattern emerges across domains, mirroring the annotation differences observed above.
\uline{Low-prior documents tend toward the micro: raw data, practical fixes, technical niches, and unstructured fragments; high-prior documents tend toward the macro: structured explanations, polished encyclopedic entries, and comprehensive analysis.}

\medskip

Analyses in \S\ref{sec:drivers-annotation} and \S\ref{sec:drivers-priors} point to a consistent pattern and answer \textbf{RQ3}:
\begin{description}[wide, itemindent=\labelsep, itemsep=\parskip, parsep=0pt, topsep=\parskip]
  \item[\textbf{RQ3.1}] Annotation protocols seem to favor comprehensive, explanatory documents on mainstream topics, leaving niche, technical, or fragmentary content unjudged.
  \item[\textbf{RQ3.2}] Retrievers exhibit relevance priors that mirror annotation biases, and rank documents resembling the relevant class higher, favoring broad, comprehensive content over niche or fragmentary documents. 
\end{description}

\section{Related Work} \label{sec:related-work}

\paragraph{Document Priors in IR.}
In IR, a \emph{document prior} traditionally refers to a query-independent signal reflecting how generally useful a document is \citep{berger_information_1999}.
Such priors have been derived from citations \citep{meij_using_2007}, web link structure \citep{craswell_relevance_2005, hauff_age_2005, kamps_importance_2008}, and content-based features such as readability \citep{bendersky_quality-biased_2011} and document length \citep{kraaij_importance_2002}.
The \emph{relevance prior} we study differs in that it is implicit: learned by dense retrievers trained on relevance judgments, emerging as a side effect of training rather than a deliberate design choice. 


\paragraph{Priors in Neural Retrievers.}
Prior work has shown that classifiers trained on relevance labels can recover a query-independent signal of document quality, and that using this signal to prune low-quality passages reduces storage and compute costs with little loss in effectiveness \citep{izacard_memory_2020, fajcik_pruning_2021, chang_neural_2024}.

We ask whether such priors can also hurt performance by making genuinely relevant documents harder to find, reframing them as a potential source of bias and linking lower relevance prior directly to lower document findability.

\paragraph{Biases and Content Preferences.}
Neural retrievers exhibit a range of content preferences: they favor fluent, formally written text \citep{macavaney_abnirml_2022}, score LLM-generated passages higher than human-written ones \citep{dai_neural_2024, wang_perplexity_2024}, prefer Wikipedia-style prose over informal language \citep{cao_writing_2025}, and are sensitive to document length, evidence placement, and lexical overlap with the query \citep{fayyaz_collapse_2025}.

We argue that such preferences can arise because retrievers learn that certain properties co-occur with relevance labels, a correlation encoded in the embedding space. What looks like a stylistic preference may thus be, in part, a relevance prior.

\paragraph{Document Findability and Retrievability.}
\emph{Findability}, the ease of locating information within a corpus \citep{morville_ambient_2005, cormacain_legislative_2017}, is operationalized by \citet{sinha_findability_2023} as how high a document is ranked for queries for which it is relevant.
A related notion, \emph{retrievability} \citep{azzopardi_retrievability_2008}, measures how often a document is retrieved across a broad query sample regardless of relevance.
We use findability because it conditions on actual relevance, which better suits our question of whether relevant documents are being missed.

This line of work is largely descriptive, leaving open \emph{what} drives findability disparities. 
We address this gap by showing that relevance priors are one such driver.

\paragraph{Annotation Artifacts and Shortcut Learning.}
The relevance prior we study is a form of annotation artifact: a correlation between document features and labels that models exploit as a shortcut \citep{geirhos_shortcut_2020, dogra_shortcut_2024}. 
When certain document types are consistently labeled relevant in training data, retrievers can learn to associate those document types' textual properties with relevance independently of query content.
The result is a prior encoded in embedding space.

Such artifacts are well documented in tasks including natural language inference \citep{gururangan_annotation_2018, karimi_mahabadi_end--end_2020}, claim verification \citep{schuster_towards_2019}, paraphrase identification \citep{zhao_comi_2024}, and question answering \citep{zellers_swag_2018}. 
Shortcut learning has also been studied in cross-modal retrieval \citep{kim_exposing_2023}, but has received less attention in text retrieval.

\section{Conclusion} \label{sec:conclusion}

We showed that supervised bi-encoders implicitly encode a document-level relevance prior (\S\ref{sec:priors-general-domain}): a query-independent signal encoded in the embedding space that reflects annotation biases rather than true relevance.
Dataset construction protocols favor comprehensive, explanatory, and self-contained documents on mainstream topics while leaving niche, fragmentary, or technical content unjudged (\S\ref{sec:drivers}).
This bias is stable enough to be picked up by retrievers as a reusable prior, associated with low-prior documents being consistently harder to retrieve even when genuinely relevant (\S\ref{sec:findability}).

These learned associations are not always harmful: some low-prior documents, such as the poorly formatted commercial pages in MSMARCO, are indeed unlikely to be relevant for most queries.
However, other documents, such as practical how-to guides and non-introductory paragraphs, are simply underrepresented in annotations, not less relevant.
Prior work has used this query-independent signal as a useful property for index pruning.
What has been overlooked is that \emph{the same signal can quietly penalize a whole class of genuinely relevant documents at retrieval time}.

\section*{Limitations}

The controlled toy experiment (\S\ref{sec:toy-experiment}) gives causal evidence that an injected, synthetic query-independent prior can create a findability gap; however, the evidence linking relevance priors to findability on real benchmarks does not prove a learned causal mechanism.
We find a correlation across multiple datasets, with BM25 as a strong control that shows substantially weaker and less consistent behavior than the supervised neural retrievers (Fig.~\ref{fig:prior-vs-findability}).
And even though this pattern holds after controlling for several confounders (\S\ref{sec:isolating-prior-effect}), we do not establish a causal link.
Our results on real benchmarks and retrievers thus give suggestive evidence: an association consistent with the prior contributing to the findability gap, but not a demonstrated causal effect.

We deliberately kept all tasks in the prior model's training data, both retrieval and non-retrieval, because our goal is to recover the prior of an actual, deployed retriever (\S\ref{sec:priors-general-domain}).
But this means our analysis on real models cannot separate the contribution of retrieval annotations from that of the rest of the fine-tuning data.
The comparison against the pre-trained backbone (App.~\ref{app:backbone-comparison}) shows that fine-tuning as a whole strengthens the separability between relevant and unjudged documents, but it cannot attribute this increase to supervised \emph{retrieval} labels specifically.

Our findings may not generalize beyond the setting we studied.
First, we focused exclusively on supervised bi-encoder dense retrievers.
Whether similar relevance priors arise in cross-encoders or learned sparse retrieval models is an open question.
Second, all datasets in our evaluation are in English.
Whether our findings transfer to multilingual or cross-lingual settings is an interesting direction for future work: the language of a document could itself function as a relevance prior, with retrievers systematically favoring some languages over others.
And lastly, our LLM-based explanation framework (\S\ref{sec:drivers}) relies on specific models, prompts, sample sizes, and generation settings.
We did not explore how sensitive the results are to these choices; different configurations might yield different explanations with varying degrees of accuracy, and therefore different insights about the drivers of annotation bias and relevance priors.

We only measured the prior's effect on retrieval performance, not on downstream tasks. 
How relevance priors affect the performance of end-to-end systems that rely on retrieval, such as RAG (Retrieval-Augmented Generation), is left for future work.
We also note that we did not find evidence of a decrease in \emph{aggregate} retrieval effectiveness, but rather reduced performance on a subset of documents relative to others.

Finally, our work is diagnostic: we showed that supervised retrievers encode relevance priors and that these are associated with a measurable findability gap, but we do not propose mitigation strategies or quantify their trade-offs.
A potential risk worth noting is adversarial misuse: knowledge of which document features are associated with high relevance priors could be exploited to craft documents that rank artificially high, independent of genuine relevance.
Whether and how to mitigate these effects, and whether this adversarial risk is significant in practice, are important directions for future work.

\section*{Acknowledgments}

\paragraph{Brno University of Technology}
This work has been supported by the Horizon EU programme through project ELOQUENCE, grant no. 101135916, and by the Ministry of Education, Youth and Sports of the Czech Republic through the e-INFRA CZ (ID:90254).

\paragraph{NodoIA San Francisco}
This work used computational resources from the \emph{NodoIA San Francisco} cluster (Ministry of Science and Technology of the Province of Córdoba, Argentina).

\bibliography{custom}

\appendix

\section{Motivating Toy Experiment} \label{app:toy-experiment}

We built all datasets from the LoTTE benchmark, using five topics: lifestyle, recreation, science, technology, and writing. We included both dev and test splits, and used both search and forum query types.
For relevance judgments, we kept only documents that are relevant to exactly one query, and queries linked to exactly one document. We also filtered out documents shorter than 10 words or longer than 400 words.
To create the document collection, we sampled non-relevant documents so they match the distribution of relevant documents by topic and length. The final collection has an overall positive rate of 0.01.
For evaluation with balanced priors, we sampled 1,000 relevant documents and added a text condition (prefix \texttt{[X]}) to half of them.
The remaining queries were used for training.

For each training query, we mined 20 hard negatives with BM25 over the document collection. 
We created training datasets with different condition prevalence values $M \in \{0.5, 0.9, 1.0\}$ among positive documents.
A positive document received the condition with probability $M$, while a negative document received it with probability $1 - M$.
A small number of hard-negative documents may still carry the condition, because a document that is relevant to one query (and thus may have received \texttt{[X]}) can be mined as a hard negative for a different query.

Dense retrievers were trained for 5 epochs on 2 A30 GPUs (24\,GB VRAM each), with a per-device batch size of 32, no gradient accumulation, and 7 negatives per query, using tevatron \citep{ma2025tevatron}.
If insufficient mined hard negatives were available, random corpus documents were sampled to fill the remainder.
We used cross-entropy loss over all passages in the global batch, so each query also sees in-batch negatives from other queries.
Following \citet{thakur_hard_2025}, we used a learning rate of $2 \times 10^{-5}$ for E5-small, and $1 \times 10^{-4}$ for Qwen2.5-0.5B with LoRA.
During training of the prior model, a document may appear with both positive and negative labels if it is relevant to some queries and sampled as a negative for others.

\begin{figure}[H]
  \centering
  \includegraphics[width=\columnwidth]{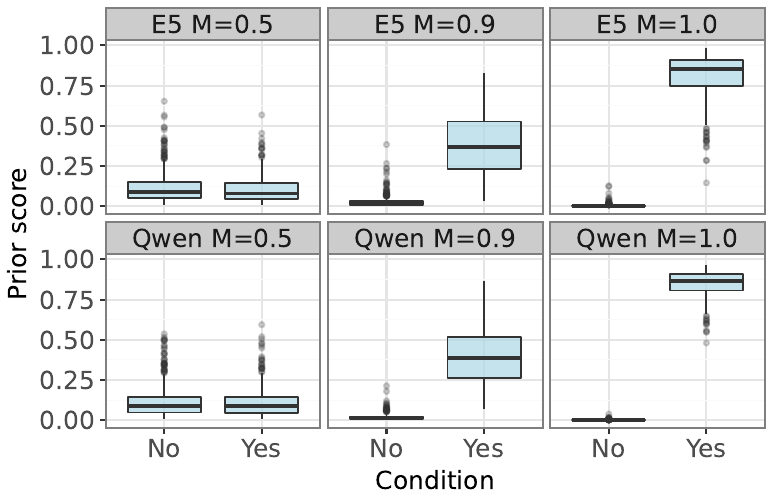}%
  \caption{
    Prior score distributions for test documents with and without the spurious \texttt{[X]} condition.
    Columns represent models trained with varying bias levels ($M \in \{0.5, 0.9, 1.0\}$).
    The top row shows the prior model trained on E5-small embeddings, while the bottom row shows the prior model trained on Qwen2.5-0.5B embeddings.
  }
  \label{fig:toy-prior-vs-condition}
\end{figure}

\begin{figure}[H]
  \centering
  \includegraphics[width=\columnwidth]{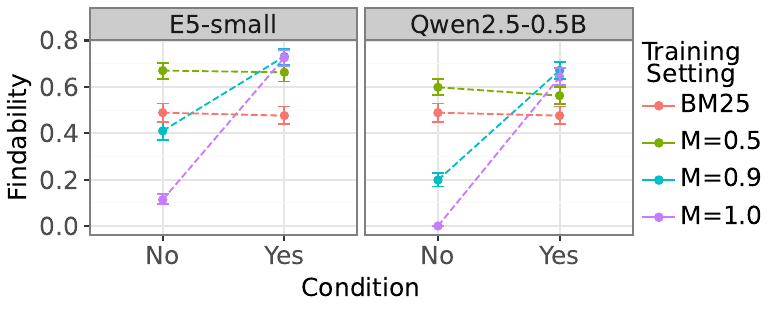}%
  \caption{
    Model findability versus the spurious \texttt{[X]} condition for test documents.
    Error bars represent 95\,\% confidence intervals over documents within each bin.
  }
  \label{fig:toy-findability-vs-condition}
\end{figure}

Figure \ref{fig:toy-prior-vs-condition} shows that retrievers trained with a stronger bias have a stronger correlation between the prior score and the condition, showing that the prior model captures the degree of bias absorbed during training.
Figure \ref{fig:toy-findability-vs-condition} shows that the findability gap between documents with and without the condition increases with the bias level $M$, confirming that relevance prior bias can directly affect retrieval performance.
BM25 shows no significant difference in findability between the two groups.

\begin{table*}[ht!]
  \small
  \centering
  \input{tables/bge_train_data_composition.tex}
  \caption{
    Composition of the training data used to fit the \textsc{bge}-based prior model.
    Positives are documents labeled relevant in \textsc{bge} training sources; negatives are mined hard negatives from the same sources.
    Some documents can appear in multiple sources, so the unique document count is lower than the sum of positives and negatives across sources.
    Moreover, some documents may appear with both positive and negative labels if they are relevant to some queries and used as negatives for others.
  }
  \label{tab:bge-train-data-composition}
\end{table*}

\begin{figure*}[ht!]
  \centering
  \includegraphics[width=1\textwidth]{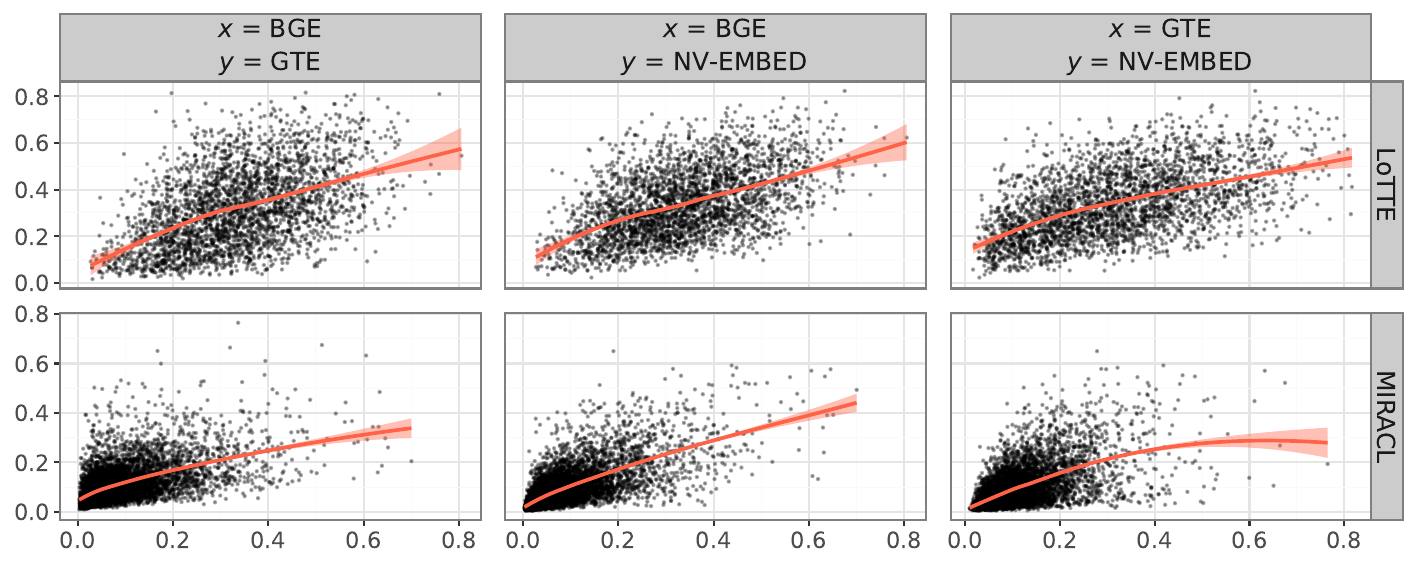}%
  \caption{
    Cross-model agreement of prior scores from models trained on different retriever embeddings.
    Each point is a document scored by two prior models.
    The red curve is a LOESS smoother with a 95\,\% confidence band.
  }
  \label{fig:prior-models-correlations}
\end{figure*}

\section{Prior Model and Findability} \label{app:prior-model}

Table~\ref{tab:bge-train-data-composition} details the composition of the training data used to fit the \textsc{bge}-based prior model.
Table~\ref{tab:prior-model-eval} reports the AUC of this prior model on held-out documents across IR datasets, quantifying how well it can separate relevant from unjudged documents based solely on the retriever's embeddings.

For evaluation, we focused exclusively on asymmetric retrieval datasets.
In these tasks, queries are typically short (e.g., questions or keywords) while documents are longer passages; unlike symmetric tasks, the roles of query and document are not interchangeable.
We used LoTTE \citep{santhanam_colbertv2_2022}, English MIRACL \citep{zhang_miracl_2023}, and selected datasets from BEIR \citep{thakur_beir_2021}, excluding any symmetric retrieval tasks and non-public datasets.
HotpotQA \citep{yang_hotpotqa_2018} involves multi-hop retrieval, where some documents only become relevant after others are found.
This means not all documents are equally relevant, making it difficult to filter documents by relevance, which is necessary for our subsequent analyses.
We therefore excluded it from our evaluation.
For BEIR, we used only the test and development splits and made sure there was no overlap with data used to train the prior model.

Figure \ref{fig:prior-models-correlations} shows scatter plots of prior scores derived from different retrievers, that is, the underlying data points used to compute the correlations in Table~\ref{tab:prior-models-correlations}.

Table~\ref{tab:findability-curves-counts} reports the number of queries and documents across datasets used for the findability analysis in \S\ref{sec:findability}.
Embedding inference for the three dense retrieval models was run on 2 A5000 GPUs (24\,GB VRAM each).
All datasets and models are publicly available and used for research purposes consistent with their intended use.
Pre-trained models were obtained from HuggingFace under their respective licenses; the OpenAI and Google Gemini APIs were used under their standard terms of service.

\begin{table}[H]
  \small
  \centering
  \input{tables/prior_model_eval.tex}
  \caption{
    Evaluation of the \textsc{bge}-based prior model on held-out documents across datasets.
    Values in the AUC column indicate 95\,\% confidence intervals computed via bootstrapping.
  }
  \label{tab:prior-model-eval}
\end{table}

\begin{table}[H]
  \small
  \centering
  \input{tables/findability_curves_counts.tex}

  \caption{
    Counts of queries and documents across datasets used for the findability analysis in \S\ref{sec:findability}.
  }
  \label{tab:findability-curves-counts}
\end{table}

\section{Priors in Fine-tuned vs.\ Pre-trained Embeddings} \label{app:backbone-comparison}

\begin{figure}[H]
  \centering
  \includegraphics[width=1.0\columnwidth]{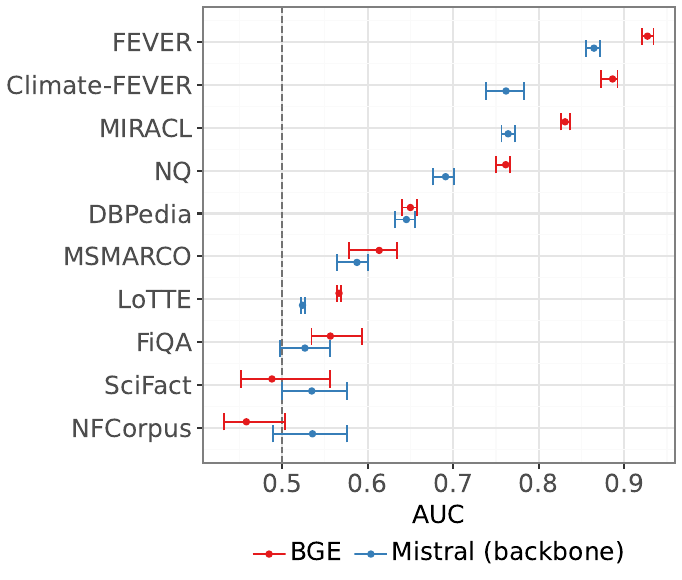}%
  \caption{
    AUC of prior models on held-out documents across datasets, comparing \textsc{bge} document embeddings against embeddings from its pre-trained backbone, Mistral-7B-v0.3.
    Error bars show 95\,\% bootstrap confidence intervals.
  }
  \label{fig:backbone-vs-finetuned-auc}
\end{figure}

We compared a prior model trained on \textsc{bge} embeddings with one trained on its pre-trained backbone, Mistral-7B-v0.3\footnote{\href{https://huggingface.co/mistralai/Mistral-7B-v0.3}{mistralai/Mistral-7B-v0.3}}. 
We randomly sampled one million documents from \textsc{bge}'s training data, the same documents used in \S\ref{sec:cross-model-consistency}, and trained one prior model using \textsc{bge} embeddings and another using Mistral embeddings, obtained via last-token pooling. 
We used the same evaluation procedure as in \S\ref{sec:priors-general-domain}.
\textsc{bge} achieved a higher AUC than Mistral on 8 out of 10 datasets, with non-overlapping 95\% intervals on 5 (Fig.~\ref{fig:backbone-vs-finetuned-auc}).
In the datasets with the highest AUC in Fig.~\ref{fig:heldout-auc} (Climate-FEVER, FEVER, MIRACL, and NQ), the gap was the largest, suggesting that fine-tuning strengthens the prior most where it is already strong. 
In the datasets where Mistral's AUC is higher than \textsc{bge}'s (NFCorpus and SciFact), both models are close to chance and have overlapping intervals.
These results further suggest that separability is induced by supervised fine-tuning, while some of the signal is already present in the pre-trained backbone. 
However, this comparison does not isolate supervised \emph{retrieval} labels from the other fine-tuning data as the source of this effect.

\section{Isolating the Prior Effect with Matched Comparisons} \label{app:isolating-prior-effect}

Table~\ref{tab:length-vs-prior} shows the correlation between document length and relevance prior across datasets: longer documents tend to have higher priors.

To isolate the effect of relevance prior on findability, we built matched document pairs $(d^H, d^L)$ where $d^H$ is a high-prior document and $d^L$ is a low-prior document that is otherwise as similar as possible.
We used Wikipedia paragraphs from the MIRACL collection, defining high-prior documents as those in the top 10\,\% of the relevance prior distribution and low-prior documents as those in the bottom 10\,\%.
We sampled 1,000 high-prior documents and matched each to the closest low-prior document by Euclidean distance over a set of ten standardized features designed to capture potential confounders, namely:
\begin{itemize}[wide, itemindent=\labelsep, itemsep=\parskip, parsep=0pt, topsep=\parskip]
  \item Named entity and numeric density (5 features): the per-token frequency of each of five word types (person, location, organization, miscellaneous, and numbers), as tagged by a NER model\footnote{\href{https://huggingface.co/dslim/bert-base-NER}{dslim/bert-base-NER}} and a PoS tagger\footnote{\href{https://huggingface.co/vblagoje/bert-english-uncased-finetuned-pos}{vblagoje/bert-english-uncased-finetuned-pos}}.
  \item Lexical complexity and diversity (3 features): the log-perplexity of the passage under Mistral-7B (\textsc{bge}'s base language model), the average inverse document frequency of content words (stemmed, stopwords excluded), and their type-token ratio.
  \item Length (1 feature): the number of content words in the passage.
  \item Style (1 feature): the log-position of the paragraph within the article
\end{itemize}
Together, these features capture document-level properties that are not strongly correlated with one another and that may independently affect retrieval difficulty.

For each document in every matched pair, we generated four synthetic queries using a two-stage prompting procedure adapted from \citet{lawrie_generate-distill_2025}, with \texttt{gpt-4.1-mini-2025-04-14}\footnote{\href{https://platform.openai.com/docs/models/gpt-4.1-mini}{platform.openai.com/docs/models/gpt-4.1-mini}} at temperature $0.3$.
In the first stage, the prompt in Table~\ref{tab:query-gen-prompt1} is applied to each passage to produce three query types: a \textit{summary} query, a \textit{rewrite} query, and a \textit{question} query.
In the second stage, the \textit{summary} and \textit{rewrite} outputs are passed to the prompt in Table~\ref{tab:query-gen-prompt2} to produce a fourth \textit{search} query, yielding a total of four queries per document.

Table~\ref{tab:matched-pair-example} presents an example pair of a high-prior and a low-prior document that are otherwise similar across all matching features, along with the queries generated for each document.
For each matched pair $(d^H_i, d^L_i)$, we computed the findability of each document over its four associated queries and computed the difference $f(d^H_i) - f(d^L_i)$.
A positive difference indicates that the high-prior document was easier to retrieve than its matched low-prior counterpart.
Aggregate results across all pairs are reported in Fig.~\ref{fig:matched-findability} in the main text.

\begin{table}[H]
  \small
  \centering
  \input{tables/length_vs_prior_correlations.tex}
  \caption{
    Pearson correlation between document length (number of whitespace-separated strings) and relevance prior across datasets.
    \\$^{***}$ p < 0.001, $^{**}$ p < 0.01, $^*$ p < 0.05.
  }
  \label{tab:length-vs-prior}
\end{table}

\begin{table}[H]
  \small
  \begin{tcolorbox}[colback=white, colframe=black, sharp corners, boxrule=0.5pt]
    \begin{flushleft}
      \input{tables/query_gen_prompt1.tex}
    \end{flushleft}
  \end{tcolorbox}
  \caption{
    First-stage query generation prompt, applied to each passage to produce \textit{summary}, \textit{rewrite}, and \textit{question} queries.
  }
  \label{tab:query-gen-prompt1}
\end{table}

\begin{table}[H]
  \small
  \begin{tcolorbox}[colback=white, colframe=black, sharp corners, boxrule=0.5pt]
    \begin{flushleft}
      \input{tables/query_gen_prompt2.tex}
    \end{flushleft}
  \end{tcolorbox}
  \caption{
    Second-stage query generation prompt, applied to the \textit{summary} and \textit{rewrite} outputs of the first stage to produce a \textit{search} query.
  }
  \label{tab:query-gen-prompt2}
\end{table}

\begin{table*}[t]
  \small
  \centering
  \input{tables/matched_pair_example.tex}
  \caption{
    Example of a matched document pair and generated queries.
    A high-prior document is paired with a low-prior counterpart that shares highly similar confounding features.
    Document texts are truncated for readability and presented alongside their synthetic queries.
  }
  \label{tab:matched-pair-example}
\end{table*}

\section{LLM-Based Explanations} \label{app:llm-explanations}

Explanations were generated using $\mathcal{M}_1$ = \texttt{gemini-3.1-pro-preview}\footnote{\href{https://ai.google.dev/gemini-api/docs/models/gemini-3.1-pro-preview}{ai.google.dev/gemini-api/docs/models/gemini-3.1-pro-preview}} at temperature $0.3$ and medium thinking level.
For each analysis, we sampled $n = 500$ documents from each group and arranged them into pairs $\mathcal{P}$.
To avoid biasing the model toward the actual labels, we referred to the low-prior or unjudged group as ``Class X'' and the high-prior or relevant group as ``Class Y''.
We then prompted $\mathcal{M}_1$ to generate an explanation $e$ of the differences between the two groups using the prompt in Table~\ref{tab:explanation-generation-prompt}.

Explanations were evaluated out-of-sample with $\mathcal{M}_2$ = \texttt{gpt-4.1-mini-2025-04-14} at temperature=0.3.
For each dataset, we sampled a balanced held-out set $\mathcal{H}$ of $m = 250$ documents ($125$ per class, $\mathcal{P} \cap \mathcal{H} = \varnothing$) and prompted $\mathcal{M}_2$ to classify each document independently using $e$ (prompt in Table~\ref{tab:explanation-evaluation-prompt}).
Classification accuracy $\text{Acc}(e)$ over $\mathcal{H}$ serves as a proxy for explanation reliability.
The balanced construction of $\mathcal{H}$ ensures $\text{Acc}(e) = 0.5$ under a random classifier.

Table~\ref{tab:labels-examples} shows examples of relevant and unjudged documents per dataset.
Tables \ref{tab:fever-climatefever-labels-explanation} and \ref{tab:miracl-nq-labels-explanation} show the explanation generated for FEVER and Climate-FEVER, and MIRACL and Natural Questions datasets, respectively, comparing relevant documents against unjudged documents.

\begin{figure}
  \centering
  \includegraphics[width=0.8\columnwidth]{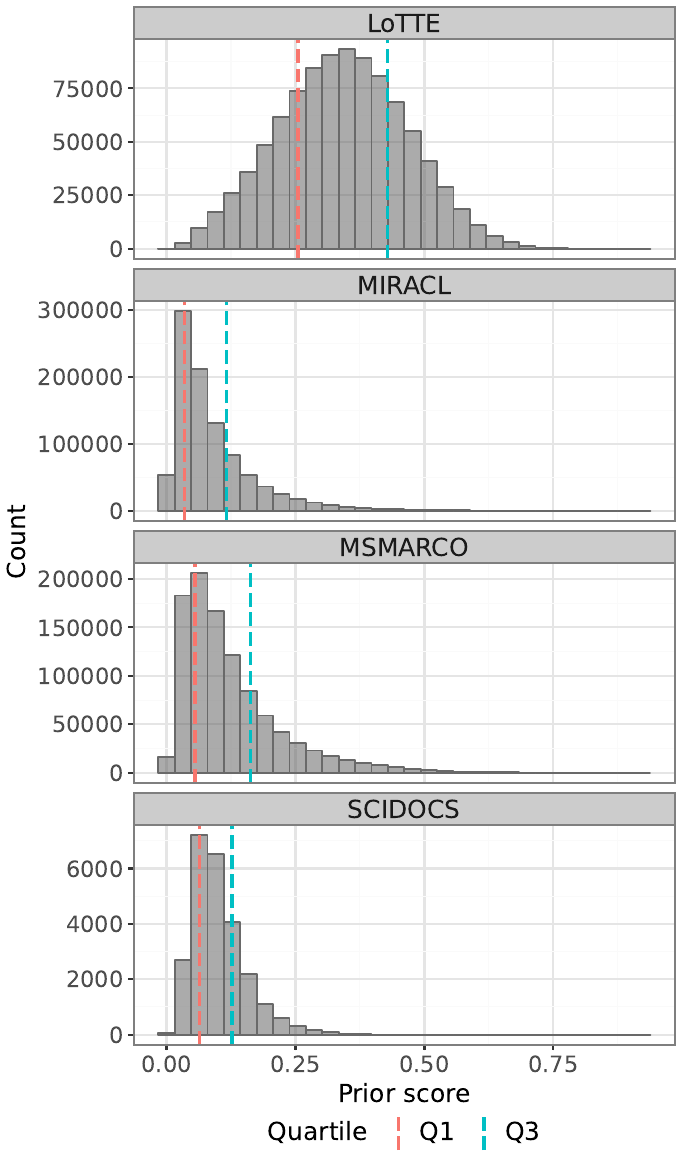}%
  \caption{
    Distribution of relevance prior scores across datasets.
    Vertical lines indicate the 25th and 75th percentiles, which are used as thresholds for defining low-prior and high-prior documents in the analysis of \S\ref{sec:drivers}.
  }
  \label{fig:prior-distribution}
\end{figure}

The 25th and 75th percentile thresholds used to define low-prior and high-prior document groups vary across datasets, reflecting differences in how priors are distributed by domain (Fig.~\ref{fig:prior-distribution}).
Table~\ref{tab:prior-examples} shows examples of high and low prior documents per dataset.
Tables \ref{tab:miracl-lotte-explanation} and \ref{tab:msmarco-scidocs-explanation} show the explanations generated for the MIRACL and LoTTE, and MSMARCO and SCIDOCS collections, respectively, comparing high-prior against low-prior documents.

\begin{table}
  \small
  \begin{tcolorbox}[colback=white, colframe=black, sharp corners, boxrule=0.5pt]
    \begin{flushleft}
      \input{tables/explanation_generation_prompt.tex}
    \end{flushleft}
  \end{tcolorbox}
  \caption{
    Prompt used for LLM-based explanation generation ($\mathcal{M}_1$).
    Class labels are anonymized as ``Class X'' (low-prior or unjudged) and ``Class Y'' (high-prior or relevant) to avoid biasing the model.
  }
  \label{tab:explanation-generation-prompt}
\end{table}

\begin{table}
  \small
  \begin{tcolorbox}[colback=white, colframe=black, sharp corners, boxrule=0.5pt]
    \begin{flushleft}
      \input{tables/explanation_evaluation_prompt.tex}
    \end{flushleft}
  \end{tcolorbox}
  \caption{
    Prompt used for out-of-sample evaluation of LLM-generated explanations ($\mathcal{M}_2$).
    The model is asked to classify held-out documents in $\mathcal{H}$ as Class X or Class Y based on the explanation $e$ derived from the first stage.
  }
  \label{tab:explanation-evaluation-prompt}
\end{table}

\begin{table*}
  \small
  \centering
  \input{tables/relevant_unjudged_examples.tex}
  \caption{
    Examples of relevant vs.\ unjudged documents for FEVER, Climate-FEVER, MIRACL, and Natural Questions.
    Documents were truncated at 250 characters to improve readability.
  }
  \label{tab:labels-examples}
\end{table*}

\begin{table*}
  \small
  \centering
  \input{tables/low_high_prior_examples.tex}
  \caption{
    Examples of documents with low vs.\ high relevance prior for MIRACL, LoTTE, MSMARCO, and SCIDOCS. Documents were truncated at 250 characters to improve readability.
  }
  \label{tab:prior-examples}
\end{table*}

\begin{table*}
  \centering
  \begin{minipage}[t]{0.48\textwidth}
    \textbf{FEVER}
    \scriptsize
    \input{tables/explanation_fever_labels.tex}
  \end{minipage}%
  \hfill
  \begin{minipage}[t]{0.48\textwidth}
    \textbf{Climate-FEVER}
    \scriptsize
    \input{tables/explanation_climatefever_labels.tex}
  \end{minipage}
  \caption{
    Explanations generated for relevant vs.\ unjudged documents in the FEVER (left) and Climate-FEVER (right) datasets.
  }
  \label{tab:fever-climatefever-labels-explanation}
\end{table*}

\begin{table*}
  \centering
  \begin{minipage}[t]{0.48\textwidth}
    \textbf{MIRACL}
    \scriptsize
    \input{tables/explanation_miracl_labels.tex}
  \end{minipage}%
  \hfill
  \begin{minipage}[t]{0.48\textwidth}
    \textbf{Natural Questions}
    \scriptsize
    \input{tables/explanation_nq_labels.tex}
  \end{minipage}
  \caption{
    Explanations generated for relevant vs.\ unjudged documents in the MIRACL (left) and Natural Questions (right) datasets.
  }
  \label{tab:miracl-nq-labels-explanation}
\end{table*}

\begin{table*}
  \centering
  \begin{minipage}[t]{0.48\textwidth}
    \textbf{MIRACL}
    \scriptsize
    \input{tables/explanation_miracl.tex}
  \end{minipage}%
  \hfill
  \begin{minipage}[t]{0.48\textwidth}
    \textbf{LoTTE}
    \scriptsize
    \input{tables/explanation_lotte.tex}
  \end{minipage}
  \caption{
    Explanation generated for low-prior vs.\ high-prior documents in the MIRACL (left) and LoTTE (right) collections.
  }
  \label{tab:miracl-lotte-explanation}
\end{table*}

\begin{table*}
  \centering
  \begin{minipage}[t]{0.48\textwidth}
    \textbf{MSMARCO}
    \scriptsize
    \input{tables/explanation_msmarco.tex}
  \end{minipage}%
  \hfill
  \begin{minipage}[t]{0.48\textwidth}
    \textbf{SCIDOCS}
    \scriptsize
    \input{tables/explanation_scidocs.tex}
  \end{minipage}
  \caption{
    Explanation generated for low-prior vs.\ high-prior documents in the MSMARCO (left) and SCIDOCS (right) collections.
  }
  \label{tab:msmarco-scidocs-explanation}
\end{table*}

\end{document}

%% file: tables/prior_models_correlations.tex
{\renewcommand{\tabularxcolumn}[1]{m{#1}}%
\renewcommand{\arraystretch}{1.15}%
\begin{tabularx}{\columnwidth}{Y C{1.5cm} C{1.5cm}}
    \toprule
    Model pair              & LoTTE & MIRACL \\
    \midrule
    \rowcolor{black!4}
    \textsc{bge} vs \textsc{gte} & 0.487 & 0.545 \\
    \textsc{bge} vs \textsc{nv-embed} & 0.507 & 0.690 \\
    \rowcolor{black!4}
    \textsc{nv-embed} vs \textsc{gte} & 0.539 & 0.584 \\
    \bottomrule
\end{tabularx}
}

%% file: tables/bge_train_data_composition.tex
{\renewcommand{\tabularxcolumn}[1]{m{#1}}%
\renewcommand{\arraystretch}{1.15}%
\begin{tabularx}{\textwidth}{Y C{2cm} C{2cm}}
\toprule
Dataset & Unique positive docs. & Unique negative docs. \\
\midrule
\rowcolor{black!4}
trivial & 506,565 & 2,433,176 \\
msmarco\_passage & 499,996 & 5,403,188 \\
\rowcolor{black!4}
eli5 & 325,293 & 316,059 \\
msmarco\_document & 314,590 & 573,564 \\
\rowcolor{black!4}
nli & 252,397 & 259,520 \\
Clustering
 (twenty\_news\_groups, biorxiv\_abstract, medrxiv\_abstract, arxiv\_title, arXiv\_abstract, biorxiv\_title, medrxiv\_title, stack\_exchangeP2P, reddit, redditP2P, stack\_exchange) & 115,475 & 674,518 \\
\rowcolor{black!4}
hotpotqa & 100,848 & 397,018 \\
quora & 51,479 & 315,207 \\
\rowcolor{black!4}
nq & 40,467 & 2,536,319 \\
stack\_overflow\_dup\_questions & 21,312 & 341,852 \\
\rowcolor{black!4}
squad & 18,891 & 18,891 \\
fiqa & 14,131 & 26,937 \\
\rowcolor{black!4}
scidocsrr & 12,793 & 81,361 \\
fever & 7,230 & 84,985 \\
\rowcolor{black!4}
arguana & 3,998 & 8,921 \\
sts & 2,286 & 13,087 \\
\rowcolor{black!4}
Classification
 (toxic\_conversations, amazon\_counterfactual, emotion, banking, amazon\_reviews, mtop\_intent, tweet\_sentiment\_extraction, imdb) & 209 & 209 \\
\midrule
\textbf{Total} & \textbf{2,280,261} & \textbf{12,981,968} \\
\bottomrule
\end{tabularx}
}

%% file: tables/prior_model_eval.tex
{\renewcommand{\tabularxcolumn}[1]{m{#1}}%
\renewcommand{\arraystretch}{1.15}%
\begin{tabularx}{\columnwidth}{Y C{1cm} C{1cm} C{1.8cm}}
    \toprule
    Dataset        & Positive docs. & Negative docs. & AUC            \\
    \midrule
    \rowcolor{black!4}
    \rowcolor{black!4}
    FEVER & 2,190 & 2,190 & (0.926, 0.940) \\
    Climate-FEVER & 1,148 & 1,148 & (0.886, 0.911) \\
    \rowcolor{black!4}
    MIRACL & 9,735 & 9,735 & (0.834, 0.845) \\
    NQ & 4,140 & 4,140 & (0.753, 0.772) \\
    \rowcolor{black!4}
    DBPedia & 6,627 & 6,627 & (0.636, 0.655) \\
    MSMARCO & 695 & 695 & (0.590, 0.648) \\
    \rowcolor{black!4}
    FiQA & 565 & 565 & (0.536, 0.603) \\
    LoTTE & 142,881 & 142,881 & (0.546, 0.550) \\
    \rowcolor{black!4}
    SciFact & 283 & 283 & (0.450, 0.545) \\
    NFCorpus & 929 & 294 & (0.426, 0.500) \\
    \bottomrule
\end{tabularx}
}

%% file: tables/findability_curves_counts.tex
{\renewcommand{\tabularxcolumn}[1]{m{#1}}%
\renewcommand{\arraystretch}{1.15}%
\begin{tabularx}{\columnwidth}{Y C{1.2cm} C{1.2cm} C{1.2cm}}
    \toprule
    Dataset        & Relevant docs. & Queries & Docs. per bin \\
    \midrule
    \rowcolor{black!4}
    Climate-FEVER & 1,148 & 1,499 & 382.7 \\
    DBPedia & 15,829 & 467 & 5,276.3 \\
    \rowcolor{black!4}
    FEVER & 2,190 & 10,923 & 730.0 \\
    FiQA & 565 & 343 & 188.3 \\
    \rowcolor{black!4}
    LoTTE & 142,884 & 26,919 & 47,628.0 \\
    MIRACL & 9,735 & 3,662 & 3,245.0 \\
    \rowcolor{black!4}
    MSMARCO & 11,521 & 7,020 & 3,840.3 \\
    NFCorpus & 3,338 & 647 & 1,112.7 \\
    \rowcolor{black!4}
    NQ & 4,140 & 3,409 & 1,380.0 \\
    SCIDOCS & 4,020 & 1,000 & 1,340.0 \\
    \rowcolor{black!4}
    SciFact & 283 & 300 & 94.3 \\
    \bottomrule
\end{tabularx}
}

%% file: tables/length_vs_prior_correlations.tex
{\renewcommand{\tabularxcolumn}[1]{m{#1}}%
\renewcommand{\arraystretch}{1.15}%
\begin{tabularx}{0.6\columnwidth}{Y C{2cm}}
    \toprule
    Dataset        & Correlation     \\
    \midrule
    \rowcolor{black!4}
    Climate-FEVER & $0.447\,^{***}$ \\
    DBPedia & $0.326\,^{***}$ \\
    \rowcolor{black!4}
    FEVER & $0.369\,^{***}$ \\
    FiQA & $0.157\,^{***}$ \\
    \rowcolor{black!4}
    LoTTE & $0.090\,^{***}$ \\
    MIRACL & $0.169\,^{***}$ \\
    \rowcolor{black!4}
    MSMARCO & $-0.011\,^{*}$ \\
    NFCorpus & $0.297\,^{***}$ \\
    \rowcolor{black!4}
    NQ & $0.171\,^{***}$ \\
    SCIDOCS & $0.319\,^{***}$ \\
    \rowcolor{black!4}
    SciFact & $0.407\,^{***}$ \\
    \bottomrule
\end{tabularx}
}

%% file: tables/query_gen_prompt1.tex
\texttt{
    \\Given the document <doc>:
    \\Step 1: produce a summary of the main concept in ten words.
    \\Step 2: rewrite the summary as a more general statement.
    \\Step 3: make a short question based on the general statement.
    \\Output the response as a json object using the format {"summary": <summary>, "rewrite": <rewrite>, "question": <question>}
}

%% file: tables/query_gen_prompt2.tex
\texttt{
    \\Given the following summaries of a document, <summary1> and <summary2>, write a search query that begins with "Find information about" such that the summarized document is relevant to me in 10 or fewer words.
    \\<summary1>
    \\\textcolor{black}{\{\{summary\}\}}
    \\</summary1>
    \\<summary2>
    \\\textcolor{black}{\{\{rewrite\}\}}
    \\</summary2>
}

%% file: tables/matched_pair_example.tex
\begingroup
\renewcommand{\tabularxcolumn}[1]{m{#1}} 
\begin{tabularx}{\textwidth}{@{} l X X @{}}
\toprule
& \textbf{High-Prior Document} & \textbf{Low-Prior Document} \\
\midrule
\textbf{Text} &
National Railway Museum. The National Railway Museum (NRM) is a museum in York forming part of the British Science Museum Group of National Museums and telling the story of rail transport in Britain and its impact on society. It has won many awards, including the European Museum of the Year Award in 2001. It is the home of the national collection of historically significant railway vehicles, as well as a collection of other artefacts and both written and pictorial records. The National Railway Museum in York displays [...] &
New Providence Building Association Stores. The New Providence Building Association Stores is a historic building located in New Providence, Iowa, United States. Numerous businesses in the town that were located in wooden structures in the central business district were destroyed by a fire on December 30, 1910. As a result, the citizens of the town banded together and formed the New Providence Building Association. It was established to buy land and construct buildings in which businesses would lease space. [...] \\
\midrule
Per. density & \multicolumn{1}{c}{0.000} & \multicolumn{1}{c}{0.000} \\
\rowcolor{black!4} Loc. density & \multicolumn{1}{c}{0.037} & \multicolumn{1}{c}{0.027} \\
Org. density & \multicolumn{1}{c}{0.037} & \multicolumn{1}{c}{0.033} \\
\rowcolor{black!4} Misc. density & \multicolumn{1}{c}{0.006} & \multicolumn{1}{c}{0.007} \\
Num. density & \multicolumn{1}{c}{0.025} & \multicolumn{1}{c}{0.027} \\
\rowcolor{black!4} Log-perplexity & \multicolumn{1}{c}{2.468} & \multicolumn{1}{c}{2.606} \\
Avg. IDF & \multicolumn{1}{c}{4.287} & \multicolumn{1}{c}{4.224} \\
\rowcolor{black!4} Type-token ratio & \multicolumn{1}{c}{0.745} & \multicolumn{1}{c}{0.740} \\
Content words & \multicolumn{1}{c}{106} & \multicolumn{1}{c}{104} \\
\rowcolor{black!4} Log-position & \multicolumn{1}{c}{0.000} & \multicolumn{1}{c}{0.000} \\
\midrule
\makecell[l]{\textbf{Synthetic} \\ \textbf{Queries}} &
\emph{[summary]} National Railway Museum in York showcases British rail transport history. \newline
\emph{[rewrite]} Museums often preserve and display historical transportation artifacts and records. \newline
\emph{[question]} Why do museums preserve and display historical transportation artifacts? \newline
\emph{[search]} Find information about British railway history and transportation museums. &
\emph{[summary]} Community built brick commercial building after 1910 fire in Iowa town. \newline
\emph{[rewrite]} Communities often rebuild commercial structures after disasters to ensure {\scriptsize economic stability.} \newline
\emph{[question]} How do communities rebuild commercial areas after disasters to maintain economic stability? \newline
\emph{[search]} Find information about community rebuilding commercial buildings after Iowa fire. \\
\bottomrule
\end{tabularx}
\endgroup

%% file: tables/explanation_generation_prompt.tex
\texttt{
    \\You will receive a list of document pairs. Each pair has one document from Class X and one from Class Y.
    \\Please read through all the pairs and then tell me:
    \\1. What are the main features of Class X documents?
    \\2. What are the main features of Class Y documents?
    \\3. What do they have in common?
    \\4. What are the main differences?
    \\Keep your answer clear and well-organized. The goal is to help someone distinguish between Class X and Class Y documents based on their characteristics.
}

%% file: tables/explanation_evaluation_prompt.tex
\texttt{
    \\Your task is to classify the given document into one of two categories: Class X or Class Y.
    \\\#\# Criteria for Classification
    \\\{\{explanation\}\}\}
    \\\#\# Output format:
    \\Reasoning: {[}briefly explain your step-by-step reasoning{]}
    \\Classification: {[}X or Y{]}
}

%% file: tables/relevant_unjudged_examples.tex
\begin{tblr}{
  colspec  = {Q[c,m,wd=1.5cm] Q[l,m,wd=0.4\textwidth] Q[l,m,wd=0.4\textwidth]},
  cell{2}{1} = {r=2}{halign=c, valign=m},
  cell{2}{2} = {bg=black!4},
  cell{2}{3} = {bg=black!4},
  cell{4}{1} = {r=2}{halign=c, valign=m},
  cell{4}{2} = {bg=black!4},
  cell{4}{3} = {bg=black!4},
  cell{6}{1} = {r=2}{halign=c, valign=m},
  cell{6}{2} = {bg=black!4},
  cell{6}{3} = {bg=black!4},
  cell{8}{1} = {r=2}{halign=c, valign=m},
  cell{8}{2} = {bg=black!4},
  cell{8}{3} = {bg=black!4},
}
\toprule
Dataset & Relevant documents & Unjudged documents \\
\midrule
  Climate-FEVER & Wind turbine A wind turbine is a device that converts the wind 's kinetic energy into electrical power .   Wind turbines are manufactured in a wide range of vertical and horizontal axis types . The smallest turbines are used for applications such as  [...] & 1953 National Challenge Cup The 1953 National Challenge Cup was the 40th edition of the USSFA 's annual open soccer championship . The Chicago Falcons defeated the Harmarville Hurricanes ( a suburban Pittsburgh team ) to win . \\
             & Solar irradiance Solar irradiance is the power per unit area received from the Sun in the form of electromagnetic radiation in the wavelength range of the measuring instrument . Irradiance may be measured in space or at the Earth 's surface after atm [...] & Urbanowo, Warmian-Masurian Voivodeship Urbanowo -LSB- urba ` nowo -RSB- ( German Zechern ) is a village in the administrative district of Gmina Dobre Miasto , within Olsztyn County , Warmian-Masurian Voivodeship , in northern Poland .   Originally in [...] \\
\midrule
  FEVER      & Ganymede (mythology) In Greek mythology , Ganymede ( -LSB- 'g\ae{}n\textsc{i},mi:d -RSB- -LSB- 'g\ae{}n\textsc{i},mid -RSB- Greek : $\Gamma$$\alpha$$\nu$$\upsilon$$\mu$$\acute{\eta}$$\delta$$\eta$$\varsigma$ , Ganymēdēs ) is a divine hero whose homeland was Troy . He was the son of Tros of Dardania , from whose name `` Troy '' was supposedl [...] & Marz Rural District Marz Rural District is a rural district ( dehestan ) in Chah Dadkhoda District , Qaleh Ganj County , Kerman Province , Iran . At the 2006 census , its population was 2,705 , in 578 families . The rural district has 38 villages . \\
             & Dog the Bounty Hunter Dog the Bounty Hunter was an American reality television series on A\&E which chronicled Duane `` Dog '' Chapman 's experiences as a bounty hunter . With a few exceptions , the series took place in Hawaii or Dog 's home state of  [...] & Oulujoki (municipality) Oulujoki ( formerly Oulun maalaiskunta or Oulu Rural Municipality ) is a former municipality of Finland . The municipality had a population of and covered a land area of 606.1 km2 . Its neighbouring municipalities were Kempele [...] \\
\midrule
  MIRACL     & Traditional English pronunciation of Latin. Latin spoken in the context of Gallo-Romance and French from approximately the 6th to the 11th-12th centuries. During this period, Latin became a primarily written language, separated from the ordinary spok [...] & Brendan Kingman. Moving into full-season ball, Kingman batted .263/.342/.372 for the Kane County Cougars. With Kevin Millar at first base, Kingman played primarily DH. Kingman moved up to the Brevard County Manatees in 1995 and hit .289\textasciitilde{}.368/.421. No [...] \\
             & Speed limits in Germany. General speed limits in Germany are set by the federal government. All limits are multiples of 5 km/h. There are two default speed limits: 50 km/h (31 mph) inside built-up areas and 100 km/h (62 mph) outside built-up areas. W [...] & Frankfort, Michigan. There were 601 households of which 18.0\% had children under the age of 18 living with them, 41.8\% were married couples living together, 10.0\% had a female householder with no husband present, 2.8\% had a male householder with no w [...] \\
\midrule
  NQ         & The Deep End of the Ocean Nine years later a young boy named Sam asks Beth if she needs the lawn mowed. Beth suspects that this boy who lives with his "father" two blocks away is in fact her lost son, and while Sam mows the lawn, she takes photograph [...] & Judge Judy On March 2, 2015, Sheindlin and CBS Television Distribution extended their contract by four years, keeping it on the air at least until completion of the 2020â€“21 season (the show's 25th).[16] \\
             & Woolly mammoth The woolly mammoth was roughly the same size as modern African elephants. Males reached shoulder heights between 2.7 and 3.4 m (8.9 and 11.2 ft) and weighed up to 6 metric tons (6.6 short tons). Females reached 2.6–2.9 m (8.5–9.5 ft) i [...] & Glenn Hoddle Hoddle prevented Swindon from slipping into the Third Division and further improvement throughout the 1991â€“92 season saw the Wiltshire club finish ninth, just missing out on a play-off place. They had briefly led the table in October. \\
\bottomrule
\end{tblr}

%% file: tables/low_high_prior_examples.tex
\begin{tblr}{
  colspec  = {Q[c,m,wd=1.5cm] Q[l,m,wd=0.4\textwidth] Q[l,m,wd=0.4\textwidth]},
  cell{2}{1} = {r=2}{halign=c, valign=m},
  cell{2}{2} = {bg=black!4},
  cell{2}{3} = {bg=black!4},
  cell{4}{1} = {r=2}{halign=c, valign=m},
  cell{4}{2} = {bg=black!4},
  cell{4}{3} = {bg=black!4},
  cell{6}{1} = {r=2}{halign=c, valign=m},
  cell{6}{2} = {bg=black!4},
  cell{6}{3} = {bg=black!4},
  cell{8}{1} = {r=2}{halign=c, valign=m},
  cell{8}{2} = {bg=black!4},
  cell{8}{3} = {bg=black!4},
}
\toprule
Dataset & High-prior documents & Low-prior documents \\
\midrule
  LoTTE      & It isn't true. Consider the toss a pair of fair, distinct, dice. Let \$A\$ be the event: "The first die comes up \$1,2\$ or \$3\$ Let \$B\$ be the event: "The first die comes up \$3,4\$ or \$5\$." Let \$C\$ be the event: " The sum of the values shown is \$9\$" Then  [...] & As another idea, you could run it so that there is no output to be placed into nohup.out: Just start the command like this: nohup command > /dev/null This will cause the output from everything to be sent to "/dev/null" and disappear. If you don't car [...] \\
             & We can model \$X\$ as a quotient space of a triangle, where two consecutive sides are glued together consistently. The third side does not get identified with anything. You can see this by thinking of a torus as a square with opposite sides identified  [...] & If the disk are mounted, and you are just looking to where they are mounted, you can type: mount That will show you which device is mounted where. If you want to see what drives are physically attached to your machine, but might not be mounted: ls -a [...] \\
\midrule
  MIRACL     & Housing in India. The most sought-after neighbourhoods of Calcutta are generally centered around Park Street, Camac Street, Lower Circular Road, Sarat Bose Road, Salt Lake, Ballygunge, Anwar Shah Road, Chowringhee and Golf Green. A recent building bo [...] & Jørn Andersen. He made his debut for Norway in 1985 and earned 27 caps, scoring five goals. His last international match was a European Championship qualifying match against Hungary in October 1990, coming on as a substitute for Jahn Ivar Jakobsen. \\
             & Bangladeshi society. Although Hindu society used to be formally stratified into caste categories, caste did not figure prominently in the Bangladeshi Hindu community. About 75 percent of the Hindus in Bangladesh belonged to the lower castes, notably  [...] & Univision Communications. In 2015, UCI signed a deal with Sling TV that grants Sling TV innovative Over-the-top programming and multi-stream rights for live and video-on-demand content from Univision Network, UniMás, UDN, Galavisión, El Rey Network,  [...] \\
\midrule
  MSMARCO    & Historians tend to argue that the anti-Nazi rhetoric of WWII helped to launch the Civil Rights Movement as it gave African Americans more of a sense that there was an opening that would allow them to demand rights. & 10 Best Laptop for Video Editing 2016. Here I find 10 best laptop for video editing 2016. Here I selecting best video editing laptops 2016 after checking all features and specs. For movie and video editing, you need good processor and graphics like I [...] \\
             & Planting varieties suited to your conditions is the best way to ensure that you'll get a healthy harvest each season. It's varieties because most pears are not self-pollinating; that is, they need at least two different trees to pollinate and produce [...] & You can type your own selector manually and restart UiExplorer from that selector by clicking Selector â" Input Selector. Continuous highlight If you click the highlight button from the toolbar you can see the selected node (if you selected just a n [...] \\
\midrule
  SCIDOCS    & Threatened egotism, narcissism, self-esteem, and direct and displaced aggression: does self-love or self-hate lead to violence? It has been widely asserted that low self-esteem causes violence, but laboratory evidence is lacking, and some contrary ob [...] & What's in an Embedding? Analyzing Word Embeddings through Multilingual Evaluation In the last two years, there has been a surge of word embedding algorithms and research on them. However, evaluation has mostly been carried out on a narrow set of task [...] \\
             & Sex differences in the human olfactory system The olfactory system (accessory) implicated in reproductive physiology and behavior in mammals is sexually dimorphic. These brain sex differences present two main characteristics: they are seen in neural  [...] & Markov Logic Networks for Optical Chemical Structure Recognition Optical chemical structure recognition is the problem of converting a bitmap image containing a chemical structure formula into a standard structured representation of the molecule. We  [...] \\
\bottomrule
\end{tblr}

%% file: tables/explanation_fever_labels.tex
\begin{Verbatim}[breaklines=true, breaksymbol={}]
**1. Main Features of Class X Documents**
* **Brevity:** The texts are extremely short, often consisting of just one to three sentences or a single brief paragraph. 
* **Niche or Obscure Subjects:** They focus on highly specific, lesser-known, or localized topics. Examples include obscure insect species (e.g., *Leucoptera andalusica*), minor geographical features (e.g., *Spatulate Ridge*), local politicians, minor sports clubs, or small villages.
* **Stub-Level Detail:** They provide only the most basic, surface-level identifying information (who, what, when, and where) without expanding on history, significance, or context. 
* **Includes Disambiguations/Lists:** Several entries are simply introductions to list pages or disambiguation pages (e.g., *James Kelley House*, *List of Tetrablemmidae species*).

**2. Main Features of Class Y Documents**
* **Depth and Length:** The texts are significantly longer, typically featuring multiple well-developed paragraphs.
* **Mainstream or Highly Notable Subjects:** They cover widely recognized, popular, and historically significant topics. Examples include famous celebrities (*Julia Roberts, Ashton Kutcher*), blockbuster movies (*Die Hard, The Book of Mormon*), major cities/regions (*Jakarta, Mississippi*), major companies (*Adidas, Microsoft Windows*), and broad academic concepts (*Physics, Dark Matter*).
* **Comprehensive Context:** Beyond basic identification, these texts delve into background, historical context, commercial or critical reception, cultural impact, and legacy.

**3. What They Have in Common**
* **Source Material:** Both classes are clearly excerpts from an encyclopedia—specifically, Wikipedia.
* **Structure:** Both classes follow the standard encyclopedic format of stating the title/subject first, followed by a definitive opening sentence that summarizes what the subject is. 
* **Tone:** Both use a formal, objective, and neutral point of view (NPOV). They avoid first-person pronouns and present information factually.

**4. Main Differences**
* **Length and Detail:** Class X documents are "stubs" (very short articles meant to be expanded later), whereas Class Y documents are robust summaries or detailed lead sections of well-developed articles.
* **Notability of the Subject:** Class Y features universally recognizable topics of global or broad cultural interest, while Class X features highly obscure topics that are only relevant to very specific fields or localities.
* **Scope of Information:** Class X tells you *what* something is and stops there. Class Y tells you what something is, *why* it matters, *how* it came to be, and its broader impact on the world.
\end{Verbatim}

%% file: tables/explanation_climatefever_labels.tex
\begin{Verbatim}[breaklines=true, breaksymbol={}]
**1. Main features of Class X documents:**
* **Brevity:** They are very short, usually consisting of just 1 to 4 sentences. They resemble "stubs" or placeholder articles.
* **Niche/Obscure Topics:** The subjects covered are highly specific, localized, or minor. Examples include lesser-known politicians, specific insect species (*Peponocranium*), minor geographical locations (a village in Poland, a local railway station in Indonesia), specific albums, and local sports clubs.
* **Superficial Detail:** They provide only the most basic identifying facts about a subject (who, what, where, and when) without delving into any deeper context, history, or significance.

**2. Main features of Class Y documents:**
* **Length and Depth:** They are much longer, typically spanning multiple well-developed paragraphs. They offer comprehensive overviews, historical context, and detailed explanations of how things work.
* **Thematic Focus (Science, Climate, and Environment):** While they cover a variety of broad topics, there is a very heavy, noticeable concentration on Earth sciences, climate change, weather phenomena, energy, and ecology (e.g., *Greenhouse effect, Sea level rise, Global cooling, Milankovitch cycles, Net metering, Tipping points in the climate system*). 
* **Broad Significance:** The topics tend to have global, historical, or scientific importance, affecting large populations or entire fields of study (e.g., *Thermodynamics, Global governance, Evolution, Economics*).

**3. What they have in common:**
* **Source Material:** Both classes are clearly excerpts from encyclopedic sources (specifically, the lead sections or introductions of Wikipedia articles).
* **Tone and Style:** Both use a formal, objective, and neutral tone appropriate for reference materials.
* **Structure:** Both almost universally begin with a definitive opening sentence that states the name of the subject and explains exactly what it is (e.g., "[Subject] is a...").

**4. Main differences:**
* **Word Count and Detail:** This is the most immediate visual difference. Class X gives the bare minimum amount of text needed to define a subject, whereas Class Y provides a miniature essay or deep-dive summary of the subject.
* **Scope of Subject Matter:** Class X deals in trivia and highly localized subjects (a specific moth, a minor 16th-century MP, a specific indie rock EP). Class Y deals with massive, overarching concepts, major world cities, and global phenomena. 
* **Thematic Consistency:** Class X is completely random across all domains of human knowledge. Class Y is heavily curated to focus on the natural environment, climate change, and the sciences.
\end{Verbatim}

%% file: tables/explanation_miracl_labels.tex
\begin{Verbatim}[breaklines=true, breaksymbol={}]
**1. Main features of Class X documents**
* **In Medias Res (Middle of the story):** These documents jump straight into highly specific details, sub-topics, or specific events related to the subject rather than introducing it. 
* **Lack of Initial Context:** They often assume the reader already knows who or what the subject is. Many begin with pronouns (e.g., "He has two daughters...", "He died at San Diego...", "It was announced...") without explicitly naming the person or thing first.
* **Narrow Scope:** The information is usually restricted to a very specific aspect of the topic, such as a person's death, a specific legal ruling, a single television season, or a family relationship.
* **Extracts from Body Paragraphs:** They read like paragraphs pulled directly from the middle or end of an encyclopedic article (such as a "Personal Life," "Death," or "History" section).

**2. Main features of Class Y documents**
* **Definitional and Introductory:** These documents generally serve as introductions or overviews. They frequently begin by explicitly defining the subject in the first sentence (e.g., *"Bucharest is the capital and largest city..."*, *"Kelp are large brown algae seaweeds..."*, *"Thomas Leo Clancy Jr. was an American novelist..."*).
* **Broad Context:** They provide a high-level summary of the topic, outlining what the subject is, its historical significance, or its main characteristics.
* **Self-Contained:** Because they establish the context immediately, a reader does not need prior knowledge of the subject to understand the paragraph. 
* **Extracts from Lead Paragraphs:** They read like the opening (lead) paragraph of an encyclopedic article.

**3. What they have in common**
* **Source Material:** Both classes consist of encyclopedic, informational text—almost certainly extracted from Wikipedia articles. 
* **Formatting:** Both begin with the title or subject name followed by a period before the text begins.
* **Tone:** Both use a formal, objective, and neutral tone appropriate for reference material.

**4. Main differences**
* **Placement in the Source Text:** Class Y represents the *beginning* (introduction/lead) of an article, whereas Class X represents the *middle or end* (body paragraphs) of an article.
* **Use of Pronouns:** Class X frequently uses unresolved pronouns (He, She, It) right at the start of the text, relying on the missing previous paragraphs for context. Class Y spells out the subject's full name or title right away.
* **Function:** Class Y is designed to *define* the subject for someone who doesn't know what it is; Class X is designed to *elaborate* on a minor or specific detail for someone who already knows the broader context.
\end{Verbatim}

%% file: tables/explanation_nq_labels.tex
\begin{Verbatim}[breaklines=true, breaksymbol={}]
### 1. Main features of Class X documents
*   **Fragmentary and Unstructured:** Class X contains many non-standard text elements, such as image captions (*"Medial surface of cerebral hemisphere. Medial view..."*), bullet points (*"• It must be written;"*), numbered list items, table notes, and raw mathematical formulas. 
*   **Context-Dependent (Dangling References):** These snippets are often pulled directly from the middle of sections or paragraphs. Because of this, they frequently rely on missing context, using unresolved pronouns or referencing things introduced earlier (e.g., *"The first two conglomerates..."*, *"signed the bill..."*, *"this era..."*).
*   **Variable Quality and Tone:** While some are well-written, many Class X documents feature poor grammar (e.g., *"is occurred from light"*), awkward phrasing, or a subjective, essay-like tone that breaks standard encyclopedia rules (e.g., *"We must recognize how interconnected..."*).

### 2. Main features of Class Y documents
*   **Coherent and Well-Structured Prose:** Class Y consists entirely of clean, complete sentences and well-formed paragraphs.
*   **Self-Contained and Introductory:** These documents are largely context-independent. They are typically the lead paragraph of an article or the introductory paragraph of a specific section (such as a Plot summary or History section). They clearly introduce and define their subjects right away.
*   **High-Quality Encyclopedic Tone:** The text strictly maintains a neutral, formal, and objective tone with proper grammar, representing polished encyclopedia entries. 

### 3. What they have in common
*   **Source Material:** Both classes are clearly excerpts sourced from Wikipedia (or a very similar online encyclopedia).
*   **Formatting:** Every document in both classes begins with the title of the article, followed by a space, and then the text snippet itself. 
*   **Markup Artifacts:** Both classes retain standard Wikipedia artifacts, including bracketed footnote citations (e.g., `[5]`, `[64]`) and editorial maintenance tags (e.g., `[citation needed]`, `[update]`, `[clarification needed]`).
*   **Topic Diversity:** Both classes cover a completely random and diverse array of topics, spanning history, science, geography, pop culture, and law.

### 4. Main differences
*   **Contextual Independence:** The most glaring difference is that **Class Y** documents make perfect sense to read on their own, whereas **Class X** documents often leave the reader confused because they jump into the middle of a thought or narrative without introducing the entities involved.
*   **Text Type:** **Class Y** is strictly standard prose (paragraphs). **Class X** is a "noisy" mix of prose, isolated lists, fragmented sentences, captions, and equations.
*   **Curation/Extraction Method:** Class Y appears to be curated to feature high-quality, high-level summaries (like article leads). Class X appears to be a random or unfiltered extraction that blindly pulls text from anywhere on the page, regardless of whether it is a caption, a list item, or a mid-paragraph sentence.
\end{Verbatim}

%% file: tables/explanation_miracl.tex
\begin{Verbatim}[breaklines=true, breaksymbol={}]
1. Main features of Class X documents
*   Short and Fragmented: They are typically very brief, often consisting of just one or two sentences, a single bullet point, or even just a section heading.
*   Highly Specific or Trivia-focused: They tend to focus on hyper-specific details, minor trivia, or raw data (e.g., specific census demographics, a single quote from a book review, or a minor sports statistic).
*   Structural Elements: Many Class X documents act as structural text for a larger page, such as introductions to lists (e.g., *"The following is the filmography of..."* or *"The first season consisted of 8 teams..."*), disclaimers, or standalone bullet points.
*   Lack of Context: Because they are so brief or specific, they usually do not provide a full understanding of the topic on their own.

2. Main features of Class Y documents
*   Longer and Well-Developed: They are generally much longer, consisting of fully fleshed-out, multi-sentence paragraphs.
*   Narrative and Descriptive: They are written in a continuous, narrative style that flows logically from one sentence to the next.
*   Comprehensive Context: They provide substantive background information, historical context, or a solid overview of the subject. They read like the introductory (lead) paragraph or a major body paragraph of an encyclopedia article.

3. What they have in common
*   Source and Tone: Both classes are clearly excerpts from an encyclopedia (specifically Wikipedia). They share a neutral, informative, and objective tone.
*   Formatting: Both classes follow the exact same formatting convention: they begin with the title or subject of the article, followed by a period, and then the text (e.g., *"Topic Name. Text goes here..."*).
*   Topic Diversity: Both classes cover a vast and identical variety of subjects, including history, geography, biology, pop culture, sports, and biographies.

4. Main differences
*   Completeness: Class Y documents are complete, self-contained thoughts that explain a topic, whereas Class X documents are often incomplete fragments, list items, or isolated facts pulled out of a larger text.
*   Length: Class Y documents are consistently longer and denser in word count compared to the brief, stub-like nature of Class X.
*   Purpose: Class Y serves to educate the reader by summarizing or describing a topic in depth. Class X serves as supplementary data, structural filler (like list introductions), or highly granular data points (like exact coordinates, specific dates of minor events, or census numbers).
\end{Verbatim}

%% file: tables/explanation_lotte.tex
\begin{Verbatim}[breaklines=true, breaksymbol={}]
### 1. Main features of Class X documents
Class X documents are highly practical, applied, and instructional. They are designed to give the reader a direct solution to a specific problem. 
* Action-Oriented & Prescriptive: They frequently use imperative verbs and tell the reader exactly what to do (e.g., *"Take a look at your syslog configuration,"* *"Put in big conduit,"* *"Write $f(z)=$..."*).
* Step-by-Step Solutions & Troubleshooting: They often provide quick fixes, code snippets, scripts, configuration paths, or direct hints to solve a roadblock (e.g., providing a bash script to monitor a process, or explaining how to fix an Ubuntu audio issue).
* Conversational & Direct Tone: The language is often informal, direct, and conversational, frequently using first- and second-person pronouns (e.g., *"I've found the process quite interesting,"* *"Hope this helps you,"* *"You can avoid shampooing your hair daily"*).
* Concise: They tend to get straight to the point, offering the "how-to" without necessarily diving deep into the exhaustive background theory.

### 2. Main features of Class Y documents
Class Y documents are highly theoretical, conceptual, and explanatory. They are designed to build the reader's foundational understanding of a topic.
* Descriptive & Analytical: Instead of telling the reader what to do, they explain *how* or *why* something works the way it does (e.g., explaining the physics of wave function collapse, how a multi-section capacitor works, or the philosophy of Schopenhauer).
* Comprehensive Proofs & Derivations: In mathematical or scientific contexts, Class Y documents tend to provide full, rigorous proofs, derivations, or detailed logical deductions rather than just a quick hint.
* Academic & Objective Tone: The language is generally more formal, pedagogical, and declarative (e.g., *"There is a difference between...,"* *"The trick to understanding why this is false is...,"* *"We know the number of divisors of..."*).
* Contextual: They often provide historical context, analogies, or deep-dives into the mechanics of a system (e.g., explaining the literary history of "Speculative Fiction" or the performance implications of RAID 10 vs. RAID 5).

### 3. What they have in common
* Source & Format: Both classes clearly originate from Q&A platforms (like Stack Exchange), forums, or technical discussion boards. 
* Goal: Both aim to answer a user's question, clarify a misunderstanding, or provide helpful information.
* Domain Diversity: Both classes cover a massive, overlapping variety of subjects, including computer science, mathematics, physics, home improvement, literature, biology, and everyday life advice.
* Technical Jargon: Both utilize domain-specific terminology and formatting, seamlessly integrating things like LaTeX for math equations, inline code blocks, or specialized vocabulary.

### 4. Main differences
* Purpose (The "How" vs. The "Why"): Class X is focused on *execution* - giving the reader the tools, commands, or steps to achieve a specific outcome. Class Y is focused on *comprehension* - giving the reader the underlying theory, proofs, or context to fully understand a concept.
* Tone: Class X reads like a helpful colleague looking over your shoulder to help you fix a bug or solve an equation. Class Y reads like a textbook, an encyclopedia entry, or a professor giving a detailed lecture.
* Content Structure: Class X relies heavily on actionable items (commands, file paths, UI navigation steps, math hints). Class Y relies on structured, logical paragraphs, conceptual analogies, and complete mathematical or scientific breakdowns.
\end{Verbatim}

%% file: tables/explanation_msmarco.tex
\begin{Verbatim}[breaklines=true, breaksymbol={}]
### 1. Main Features of Class X Documents
*   Encoding Errors (Mojibake): Class X is heavily plagued by character encoding issues. Smart quotes, apostrophes, and bullet points are frequently rendered as garbled characters (e.g., `â€™` for an apostrophe, `â€œ` and `â€ ` for quotation marks, `â€¢` for bullets, and `Â½` for fractions).
*   Parsing and Formatting Glitches: There are frequent text-extraction errors where the first letter of a sentence or paragraph is missing after a period, causing words to merge (e.g., "feet.he islands", "benefits.his is", "print.his applies", "teacher.ecome").
*   Broken Lists: When lists appear in Class X, the formatting is usually broken. You will see floating, unpunctuated numbers inserted awkwardly into the text (e.g., "1 No PNC fees", "1 Accessibility:", "1 The FULTON County GA ZIP Code Map").
*   Conversational, Commercial, or Raw Web Tone: The content often reads like raw, unpolished web scraping. It includes forum posts ("DIY: Differential Fluid Change"), classified ads ("$686 Oct 14 JUST UNLOADED"), local business listings, and promotional/marketing copy ("We believe in doing more than delivering a better insurance solution").

### 2. Main Features of Class Y Documents
*   Cleaner Text and Punctuation: Class Y documents are generally much cleaner. They use standard ASCII apostrophes (`'`) and quotation marks (`"`), avoiding the severe, distracting encoding errors seen in Class X. 
*   Factual and Encyclopedic Tone: The content reads like excerpts from encyclopedias, textbooks, dictionaries, or direct answers from a Q&A database (e.g., Wikipedia, WebMD). They are highly objective and informative.
*   Structured Formatting: When lists or steps are included, they are usually formatted properly with standard punctuation (e.g., "1. Crystallization from Magma 2. Precipitation").
*   Direct Answers: Many Class Y documents are written in a way that directly answers a specific, implicit question (e.g., defining a medical condition, explaining a historical event, or detailing a scientific process).

### 3. What They Have in Common
*   Source Material: Both classes consist of short text passages or snippets scraped from the internet (likely from a search engine dataset, such as MS MARCO).
*   Length: Documents in both classes are roughly the same length, typically ranging from 2 to 4 sentences (about 40 to 100 words).
*   Topic Diversity: Both classes cover a massive variety of subjects, including medicine, history, pop culture, geography, finance, and DIY instructions.
*   Occasional Scraping Artifacts: While much more prominent in Class X, Class Y is not entirely immune to web-scraping artifacts. Class Y occasionally features minor typos, weird spacing, or slight encoding issues with foreign accents (e.g., "brontÃ«").

### 4. Main Differences
*   Text Quality and Readability: The most immediate visual difference is cleanliness. Class X is messy, filled with `â€™` symbols, missing letters, and broken HTML list numbers. Class Y is significantly cleaner, better parsed, and easier to read.
*   Intent and Style: Class X contains a lot of subjective, commercial, or user-generated content (ads, blogs, forums, marketing pitches). Class Y is almost exclusively objective, educational, and factual.
*   Contextual Cohesion: Class Y snippets usually form a complete, cohesive thought or definition. Class X snippets often feel like random chunks of text mashed together from different parts of a webpage (e.g., mixing a headline, a price, and a random bullet point). 

Summary: If you are trying to distinguish between the two, look at the punctuation and the tone. If the text is full of `â€™` symbols, floating numbers, and reads like a forum post or an ad, it is Class X. If the text uses standard apostrophes (`'`) and reads like a clean, factual Wikipedia or dictionary entry, it is Class Y.
\end{Verbatim}

%% file: tables/explanation_scidocs.tex
\begin{Verbatim}[breaklines=true, breaksymbol={}]
### 1. Main Features of Class X Documents
*   **Domain Focus:** Almost exclusively focused on Computer Science, Engineering, Artificial Intelligence, Mathematics, and Information Technology.
*   **Research Type:** They primarily describe the invention, optimization, or evaluation of new algorithms, computational models, software systems, or hardware architectures.
*   **Formatting:** The abstracts are universally **unstructured**, consisting of a single, continuous paragraph without explicit section headers.
*   **Terminology:** Highly technical and computational. Common keywords include *optimization, framework, neural networks, baseline, latency, routing, dataset, simulation,* and *algorithm*.
*   **Evaluation:** Success is usually measured by computational efficiency, accuracy, speed, bandwidth, or error reduction compared to a baseline system.

### 2. Main Features of Class Y Documents
*   **Domain Focus:** Highly multidisciplinary, with a very strong presence of Medicine, Psychology, Public Health, Biology, and Social Sciences (though it does include some Computer Science and Business papers as well).
*   **Research Type:** They heavily feature empirical studies, clinical trials, observational research, behavioral analysis, and systematic literature reviews. 
*   **Formatting:** Many of these documents use **structured abstracts** with explicit, capitalized headings (e.g., *BACKGROUND, OBJECTIVES, METHODS, RESULTS, CONCLUSIONS*).
*   **Terminology:** Focuses on experimental design, human/biological factors, and statistical significance. Common keywords include *participants, patients, symptoms, treatment, p-values (e.g., p < 0.01), confidence intervals (CI),* and *surveys*.
*   **Evaluation:** Success or findings are usually measured by statistical correlations, health outcomes, psychological metrics, or behavioral changes in a sample population.

### 3. What They Have in Common
*   **Academic Nature:** Both classes consist of academic/scientific abstracts that summarize peer-reviewed research papers.
*   **Core Structure:** Regardless of formatting, both classes follow the standard scientific narrative arc: identifying a problem/gap in the literature, proposing a method or study to address it, and reporting the results/implications.
*   **Overlap in CS:** Both classes contain papers related to Computer Science and Machine Learning, though Class Y tends to apply these technologies to human-centric or biological problems (e.g., medical image segmentation, social media behavior).

### 4. Main Differences
*   **Discipline & Subject Matter:** Class X is strictly technical, dealing with machines, code, networks, and math. Class Y is largely human-centric or biological, dealing with patients, psychology, health, and social behaviors.
*   **Abstract Formatting:** If an abstract is broken down into capitalized sections (METHODS, RESULTS, etc.), it belongs to Class Y. Class X uses traditional, single-block paragraphs.
*   **Methodology:** Class X relies on mathematical proofs, system building, and benchmark testing against datasets. Class Y relies heavily on the scientific method applied to the physical world, utilizing control groups, human participants, clinical observations, and statistical hypothesis testing.
\end{Verbatim}